\documentclass[reviewcopy]{elsarticle}

\usepackage[reviewcopy]{adndt}
\usepackage{longtable}

\usepackage{mathptmx}

\usepackage{amsmath}

\usepackage{amssymb}

\biboptions{square,sort&compress}
\bibpunct[]{[}{]}{,}{n}{}{;}
\begin{document}

\begin{frontmatter}

\journal{Atomic Data and Nuclear Data Tables}


\title{Energy levels, radiative rates and electron impact excitation rates for transitions in  Si~III}

  \author[One]{K. M. Aggarwal\corref{cor1}}
  \ead{K.Aggarwal@qub.ac.uk}
 

  \cortext[cor1]{Corresponding author.}

  \address[One]{Astrophysics Research Centre, School of Mathematics and Physics, Queen's University Belfast,\\Belfast BT7 1NN,
Northern Ireland, UK}


\date{06/05/2016} 

\begin{abstract}  
Energy levels and radiative rates (A-values) for four types of transitions (E1, E2, M1, and M2) are reported for an astrophysically important Mg-like ion Si~III, whose emission lines have been observed in a variety of plasmas. For the calculations, well-known and widely-used {\sc grasp} code has been adopted, and results are listed  for transitions among the 141 levels of the 3$\ell3\ell'$ and 3$\ell$4$\ell$ configurations.  Experimental energies are available for only the lowest 58 levels but there is no major discrepancy with theoretical results. Similarly, the A-values and lifetimes show a satisfactory agreement with other available results, particularly for strong E1 transitions. Collision strengths are also calculated, with the {\sc darc} code, and listed for resonance transitions over a wide energy range, up to 30~Ryd. No similar results are available in the literature for comparisons. However, comparisons are made with the more important parameter, effective collision strength ($\Upsilon$), for which recent $R$-matrix results are available for a wide range of transitions, and over a large range of temperatures. To determine $\Upsilon$, resonances have been resolved in a narrow energy mesh, although these are not observed to be as important as for other ions. Unfortunately, large discrepancies in $\Upsilon$ values are noted for about half the transitions. The differences increase with increasing temperature and worsen as the upper level J increases. In most cases the earlier results are overestimated, by up to (almost) two orders of magnitude, and this conclusion is consistent with the one observed earlier for Be-like ions. \\

Received 06 October 2016; Accepted 11 November 2016 \\

{\bf Keywords:} Mg-like Si~III, energy levels, radiative rates, lifetimes, collision strengths, effective collision strengths

\end{abstract}

\end{frontmatter}




\newpage

\tableofcontents
\listofDtables
\listofDfigures
\vskip5pc


\section{Introduction}
\label{intro}
Emission lines of many Mg-like ions have been observed in a variety of astrophysical plasmas, such as solar, early and late-type stars and planetary nebulae -- see for example, \cite{dk1} and references therein. Lines from several of these ions (such as Ca~IX, Ti~XI and Fe~XV) are also prominent in fusion plasmas. However, to interpret observations and to model these plasmas atomic data are required for several parameters, including energy levels, radiative rates (A-values) and effective collision strengths ($\Upsilon$). Generally, energy levels for these ions are fairly well known, and the compilation of assessed experimental data are freely available from the NIST (National Institute of Standards and Technology)  website {\tt http://www.nist.gov/pml/data/asd.cfm}. However, corresponding informations for A- and $\Upsilon$ values are not available from measurements, but over the past few decades several workers have reported theoretical results for many of the Mg-like ions -- see for example the data stored in the CHIANTI database at {\tt http://www.chiantidatabase.org/} or references in \cite*{icft1}. Most of these data, particularly for $\Upsilon$, are limited to a few levels/transitions, and therefore require extension. More importantly, for some ions (such as P~IV, Cl~VI and K~VIII) no collisional data are available.

Realising the importance of atomic data for Mg-like ions, recently \cite{icft1} have reported calculations for a wide range of ions, up to Z = 36. They have considered a large number of levels (283 belonging to the 3$\ell3\ell'$, 3$\ell$4$\ell$ and 3$\ell$5$\ell$ configurations) and have reported a consistent set of results for energy levels, A-values and $\Upsilon$. For the determination of atomic structure they have adopted the {\em AutoStructure} (AS) code of \cite{as}, and for the collisional calculations the $R$-matrix code of \cite*{rm2}. Furthermore, they have resolved resonances in thresholds region and therefore, their data should be the best available to date.

The $R$-matrix code of \cite{rm2}, adopted by \cite{icft1}, basically calculates collision strengths ($\Omega$) in $LS$  coupling (Russell-Saunders or spin-orbit coupling), and in order to calculate $\Omega$ (and $\Upsilon$) for {\em fine-structure} transitions, they utilised their {\em intermediate coupling frame transformation} (ICFT) method \citep*{icft}. Unfortunately, in the recent literature questions have been raised about the reliability of their approach. For example, we \citep{alx1,ciii} have demonstrated that the implementation of such an approach leads to a significant overestimation (of orders of magnitude) of $\Upsilon$ values over a wide range of temperatures for a large number of transitions in Be-like ions. Similar overestimations have also been noted for transitions in Al-like Fe~XIV \citep{fe14} and Ar-like Fe~IX \citep{fe9}.

However, in a series of papers \citep*{icft2, icft3, icft4} the overestimation of $\Upsilon$ results with the ICFT methodology has been justified on the basis of the larger calculations performed by the group authors. For example, we included only 98 levels of the 2$\ell$2$\ell'$, 2$\ell$3$\ell'$ and  2$\ell$4$\ell'$  configurations for most Be-like ions \citep{alx1, ti19,cl14,alx}, and only 166 for C~III \citep{ciii}, the additional 68 levels belonging to  the 2$\ell$5$\ell'$  configurations, well short of 238 considered by \cite{belike}. However, in a recent paper \citep*{niv} we considered exactly the same 238 levels for N~IV and arrived at the same conclusion that the ICFT results for $\Upsilon$ are indeed overestimated, by up to four orders of magnitude for over 40 per cent of the transitions. Moreover, the overestimation of $\Upsilon$ results is  over the whole range of temperatures. Therefore, it has become necessary to test the (in)accuracy of the $\Upsilon$ results of \cite{icft1} for Mg-like ions.

We also note here that large discrepancies in the ICFT calculations of \cite{icft5} for transitions in Mg-like Fe~XV were observed earlier \citep{fe15} -- see also section~5 and Table~E for transitions in Si~III. The error in the code was later rectified by \cite{icft6}. Moreover, in certain circumstances the ICFT approach does lead to the overestimation in the $\Upsilon$ values, as discussed by \cite*{oiii} for transitions in O~III, and also explained by \cite{icft2}. Nevertheless, in this paper we consider the results for Si~III, which is not only an important Mg-like ion but its atomic data have recently been 'benchmarked' by \cite*{bench}.

Lines of Si~III have been extensively analysed by many workers -- see for example, \cite{dk1} and \cite{bench} and references therein. Of particular interest is the 120.7~nm emission line arising from the 1s$^2$~$^1$S$_0$ -- 3s3p~$^1$P$^o_1$ transition, which has been extensively observed in both solar and stellar plasmas -- see for example, \cite{jll}. The early close-coupling calculations for $\Upsilon$ using the $R$-matrix method were undertaken by \cite*{bbk}. They considered only 12 lowest lying $LS$ states of the 3s$^2$, 3s3p, 3s3d, 3p$^2$, 3s4s and 3s4p configurations, and reported results for $\Upsilon$ over the 5$\times$10$^3$ to 2.5$\times$10$^5$~K temperature range, sufficient for analysis of observations because the temperature of  maximum abundance in ionisation equilibrium for Si~III is only $\sim$50~000~K  \citep*{pb}. However, an error was later detected in their work and was rectified by \cite{dk1}, whose collisional data have mostly been utilised for observational analysis -- see for example \cite{bald}, and are also stored in the CHIANTI database. Nevertheless, their data remain for limited transitions among 20 fine-structure levels of the above listed 12 states, and hence are not fully sufficient for observational analysis because some of the strong lines of Si~III are associated with higher excited levels, such as 3s4f~$^1$F$^o_3$ \citep{bench}.

A much larger calculation involving 45 fine-structure levels belonging to 25 $LS$ terms  ($n \le$ 4) of four Mg-like ions, including Si~III, was later performed by \cite{icft5}, but they reported results for $\Upsilon$ for only 15 transitions from the ground 3s$^2$~$^1$S$_0$ to higher excited levels -- see their Table~3. However, their $\Upsilon$ results for all transitions are now available on the website:  {\tt http://www.open.adas.ac.uk}.   Nevertheless, since their similar results for Fe~XV were clearly demonstrated  to be inaccurate \citep{fe15}, as already stated, we will focus our comparisons with the most recent and relevant results of \cite{icft1}, discussed earlier.

As in our earlier works, we employ the  fully relativistic {\sc grasp} (General-purpose Relativistic Atomic Structure  Package) code for the generation of wavefunctions, i.e. to determine the atomic structure of Si~III. This code was  originally  developed by  \cite{grasp0}, but  has since undergone through multiple revisions. The version adopted here has been significantly revised by Dr. P. H. Norrington, one of the authors. This version is known as GRASP0 and is available at the website: {\tt http://amdpp.phys.strath.ac.uk/UK\_APAP/codes.html}.  Similarly, for the scattering calculations we have adopted the relativistic version of the $R$-matrix code, known as DARC  (the Dirac atomic  $R$-matrix code), and available at the same website. Both these codes have been adopted because of their reliability and our past experience with these for a wide range of ions. Otherwise, it is fair to state that the relativistic effects (included in these codes) are not too important for a moderately heavy Si ion. However, because of the inclusion of fine-structure in the definition of channel coupling, the size of the Hamiltonian (H) matrix increases substantially, and thus makes the calculations computationally more demanding. For this reason, our calculations include only 141 levels of the 3$\ell3\ell'$ and 3$\ell$4$\ell$ configurations, 18 in total. Therefore, our calculations are comparatively smaller than those performed by \cite{icft1}, because for practical reasons we are omitting the 142 levels of the 3$\ell$5$\ell$ configurations. Nevertheless, our results should be sufficient to draw the necessary conclusions, as were the cases with our smaller calculations for Al~X \citep{alx1} and C~III \citep{ciii}, i.e. the Be-like ions.

\section{Energy levels}
\label{calc}

Our energies obtained with  an `extended average level' (EAL) approximation are listed in Table~1 along with the experimental values compiled by NIST.  For the 141 level calculations (GRASP1) energies obtained {\em with} and {\em without} the contributions of Breit and QED (quantum electrodynamic) effects are listed, whereas for the 283 levels (GRASP2) only the final (corrected) energies are given for the comparison purpose. This is because the contribution of Breit and QED effects is almost negligible for most levels and is below 0.04~Ryd for a few, such as 43--48, see  columns under GRASP1a and GRASP1b. This is quite expected because Si~III is a moderately heavy ion.  However, the inclusion of Breit and QED effects slightly changes the ordering for a few levels, such as 79--84, 87/89 and 105--111.

Experimental energies are available for only levels below 58, and differences for a few with our (GRASP1b) energies are below 0.04~Ryd, see for example levels 16, 43 and 46--48. Similarly the orderings between theory and measurements are compatible for most levels, although there are minor differences for a few, such as 28/29 and 52/53.

For some ions, such as Si~II \citep{si2}, inclusion of additional CI (configuration interaction) appreciably affects the energy levels. Therefore, to assess its impact we have performed another calculation (GRASP2) which includes the same 283 levels as by \cite{icft1}. However, for most levels of Si~III there is no appreciable discrepancy between the GRASP1 and GRASP2 energies, although differences for a few are up to 0.02~Ryd -- see for example, 78--81. Additionally, for two levels the GRASP2 energies differ by 0.05~Ryd ($<$2\%), higher for 86 (3d$^2$~$^1$G$_4$) but lower for 141 (3d4d~$^1$S$_0$), and therefore there is no consistency. Nevertheless, the energies calculated with the GRASP code are compatible with those obtained with another independent code, i.e.  the {\em Flexible Atomic Code} ({\sc fac}) of \cite{fac} -- see energies under column FAC1 in Table~1. {\sc fac} is also a relativistic code and generally provides comparable results for energy levels. For Si~III also, the GRASP2 and FAC1 energies agree closely within 0.03~Ryd and the orderings are also nearly the same. Although this result was expected, the exercise became desirable in the absence of measurements for higher excited levels of Si~III.  For the same reason we have performed yet another calculation, i.e. FAC2, which includes much more CI with 1211 levels of the 3*2 and 3*1 $n$*1 ($n \le$ 9) configurations. Although a few levels in FAC2 (such as 43, 53 and 90) cannot be be unambiguously  identified, there is no appreciable discrepancy between the FAC1 and FAC2 energies, i.e. there is no clear advantage in including a larger CI as far as the levels of Si~III are concerned. However, this calculation leads to another important conclusion and that is the strong intermixing of levels from higher configurations with those of $n \le$ 4  --  see also the energy table of  \cite{icft1} for the levels of Si~III. As a result of this if we want to include all 141 levels listed in Table~1 in a collisional calculation then we have to include a further $\sim$500 levels, because resonances arising from the intermixed levels of higher configurations may considerably affect the calculations of $\Upsilon$ -- see section 5. However, with the computational resources available with us such a large calculation (with about $\sim$650 levels) is not feasible and therefore we had to make a compromise, although the levels listed in Table~1 are {\em not} the lowest.

Finally, in Table~1 we include the energies of \cite{icft1} calculated with the {\sc as} code, because we will be comparing our collisional data with their work. There are some minor differences in level orderings, see for example, 10--12, 43--45 and 51--52. However, there is no significant discrepancy between our GRASP2 and the AS energies (because both calculations include the same CI), and differences for a few (such as 113--115) are below 0.1~Ryd. Therefore, with all comparisons discussed above we may confidently state that determination of energy levels for Si~III is not problematic and all results listed in Table~1 are accurate to better than 2\%.

\section{Radiative rates and lifetimes}
\label{rad}
For modelling applications the most dominant and important are the A-values for electric dipole (E1) transitions. However, for a better accuracy of plasma modelling, similar A-values for  electric quadrupole (E2), magnetic dipole (M1) and  magnetic quadrupole (M2) transitions are also desired. Therefore, we have calculated A-values for all four types and note that these are related to the   f-values (oscillator strengths) as  

\begin{equation}
f_{ij} = \frac{mc}{8{\pi}^2{e^2}}{\lambda^2_{ji}} \frac{{\omega}_j}{{\omega}_i} A_{ji}
 = 1.49 \times 10^{-16} \lambda^2_{ji}  \frac{{\omega}_j}{{\omega}_i}  A_{ji} 
\end{equation}
where $m$ and $e$ are the electron mass and charge, respectively, $c$ the velocity of light,  and $\omega_i$ and $\omega_j$  the statistical weights of the lower ($i$) and upper ($j$) levels, respectively. Our calculated results, in the length  form, are listed in Table~2 for the energies/wavelengths ($\lambda$, in $\rm \AA$), radiative rates (A$_{ji}$, in s$^{-1}$), oscillator strengths (f$_{ij}$, dimensionless), and line strengths (S, in atomic unit =  6.460$\times$10$^{-36}$ cm$^2$~esu$^2$) for all  E1 transitions. However, for the E2, M1 and M2  transitions  only the A-values are listed in Table~2. Furthermore, for brevity only transitions from the lowest 29 to higher excited levels are listed in Table~2, but full table is  available online in the electronic version (see Appendix A).

Several workers in the past have calculated A-values (mainly) for E1 transitions of Mg-like ions -- see for example \cite*{saf} and reference therein. These authors have also performed large calculations for transitions among the 3$\ell$3$\ell'$ configurations of all ions with 13 $\le$ Z $\le$ 100. They have employed their relativistic {\em many-body perturbation theory} (MBPT), but have reported limited results for most ions, including Si~III. Nevertheless, \cite{kp} have compiled and critically assessed the A-values from many sources (including those from \cite{saf}) and their recommendations (of varying inaccuracy A to E or equivalently 3 to 100\%) cover the largest number of transitions, but mostly belonging to higher levels of Si~III. Among the lowest 38 levels (see Table~1)  \cite*{cff} have determined A-values  with the  {\em multi-configuration Hartree-Fock} ({\sc mchf}) code. Their results are also available  on the website: {\tt  http://nlte.nist.gov/MCHF/view.html}. In Table~A we compare our f-values with the {\sc grasp} (GRASP1 and GRASP2) and {\sc fac} (FAC1 and FAC2) codes among the lowest 22 levels. For comparisons the corresponding results with {\sc mchf} are also included. For comparatively strong transitions with large f-values ($>$ 0.1) all calculations agree within about 20\%, which is highly satisfactory. The only exception is the 5--14 (3s3p~$^1$P$^o_1$ -- 3p$^2$~$^1$S$_0$, f $\sim$ 0.3) transition for which our GRASP1, GRASP2, FAC1 and FAC2 f-values are consistent, but the {\sc mchf} result is lower by about 30\%, and has been recommended by \cite{kp}. Such anomalies for a few transitions are often found and mainly arise with differing amount of CI and/or methodology. For the same reason, variations in the f-values for weaker transitions are up to a factor of three (or even higher) for a few, such as   1--3/18/20, because the additive or cancellation effect of multiple mixing coefficients is much greater on these.

Another way to assess the accuracy of f-values is to compare the ratio (R) of the velocity and length forms. A value closer to unity generally gives an indication about the accuracy of the  results, although the length form is normally considered to be more accurate. Therefore, in Table~A we have also listed R from our GRASP1 and GRASP2 calculations, but stress here that near unit value of R is only a desirable criterion, not a necessary one because often even for strong transitions calculations with differing amount of CI may give R $\sim$ 1, but completely different results in magnitude \cite{co16}.  For almost all strong (and many weaker) transitions listed in Table~A, R is within 20\% of unity and therefore indicates about the reliability of our results listed in Table~2. 

For some E2, M1 and M2 transitions the A-values are  also available from the {\sc mchf} calculations (see also \cite{kp}) and in Table~B we make  comparisons with our GRASP1 and GRASP2 results. There is no discrepancy between the three calculations for these transitions. Finally, we compare lifetime ($\tau$ = 1.0/${\sum_{i}} A_{ji}$) in Table~C for the lowest 29 levels of Si~III. The only $\tau$ results available in the literature for the levels of Si~III are by \cite{saf} and  \cite{cff} with the  {\sc mbpt}  and {\sc mchf} codes, respectively. Their results are included  in Table~C for comparisons. From the {\sc mchf} code there are two sets of $\tau$  values, i.e. MCHF1 and MCHF2, obtained with {\em ab initio} and {\em adjusted} energies, respectively. The two sets of $\tau$ mostly agree within 20\%, but differences for two levels (22 and 23) are up to 36\%. Similarly, for most levels there is no (major) discrepancy between the GRASP and MCHF (and MBPT) results, but for a few the differences are striking. Particularly noteworthy are the 3p3d~$^3$F$^o_{2,3,4}$ (21--23) levels, because for  these the differences between the GRASP and MCHF  \,$\tau$ are up to a factor of three. These differences directly relate to the corresponding differences in A-values of the dominating E1 transitions, which are invariably weak. Unfortunately, discrepancies with the MBPT results of \cite{saf} are even larger. However, for  degenerating levels of other states (such as $^3$P and $^3$D), $\tau$ values are (nearly) the same in all calculations, but differ for the 22 and 23 levels ($^3$F$^o_{3,4}$) in the MCHF work.  

The accuracy of the A-values can (indirectly) be assessed by making comparisons with measurements of lifetimes, which are available for a few levels, listed in Table~C. Berry et al.  \cite{berry} and later  Bashkin et  al. \cite{bash} have measured $\tau$ for a few levels by beam foil experiments. For the 3p$^2$~$^1$D$_2$ level the theoretical results are higher by up to (nearly) a factor of two, but the agreement between theory and measurements is satisfactory for the other remaining levels, particularly with the later measurements of \cite{bash}. Similarly, \cite{hsk} have measured $\tau$ for the 3s3p~$^3$P$^o_1$ level to be 59.9$\pm$3.6~$\mu$s, which compares well with the MCHF work but is lower by $\sim$40\% than our or the MBPT calculations.  Finally, \cite{liva, livb} have measured $\tau$ for two levels, namely 3s4s~$^3$S$_1$ and 3p$^2$~$^1$S$_0$, but these are higher (by up to 40 per cent) than all theoretical results, listed in Table~C.  However, this limited comparison is not sufficient for accuracy assessment of the larger data reported in the paper. Moreover, our emphasis is on the collisional calculations (described in the next section) and hence the determination of atomic structure has scope for improvement.

\section{Collision strengths}

To  calculate  $\Omega$ the $R$-matrix radius adopted for  Si~III is 19.2 atomic units, and 45  continuum orbitals have been included for each channel angular momentum in the expansion of the wavefunction.  The maximum number of channels generated for a partial wave is 729, which makes  the size of the (largest) Hamiltonian (H) matrix to be 32~835. However, this large expansion allows us  to compute  $\Omega$ up to an energy of  $\sim$30~Ryd. Considering that the highest threshold is at 3.4~Ryd (see Table~1) and the temperature of  maximum abundance in ionisation equilibrium for Si~III is only $\sim$50~000~K  \citep{pb} (i.e. $\sim$ 0.32~Ryd), the energy range included in the calculations is {\em well  above} what may be required, but allows us to calculate values of effective collision strengths ($\Upsilon$) up to 1.8$\times$10$^6$~K, without any requirement for the extrapolation of energy range for $\Omega$ -- see Eq. (3).  In contrast, \cite{icft1} calculated values of $\Omega$ only up to 7.4~Ryd  (i.e. less than 4~Ryd above thresholds) but reported $\Upsilon$ values up to T$_e$ = 1.8$\times$10$^7$~K, equivalent to 114~Ryd. Therefore, they {\em extrapolated} values of $\Omega$ over a very wide energy range, and this has been a major source of inaccuracy in their results, as discussed earlier on several occasions \citep{alx1, ciii, niv}.

Furthermore,  for calculating $\Omega$ we have considered all partial waves with angular momentum $J \le$ 40.5, sufficient for  convergence  for a majority of transitions and at most energies.  However, for some allowed transitions and particularly towards the higher end of the energy range, our $J$ range is not fully sufficient for the convergence of $\Omega$. Therefore, to account for the  higher neglected partial waves,  we have included the contributions  through the Coulomb-Bethe  \citep*{ab} and  geometric series  approximations for allowed and forbidden transitions, respectively. 

The electron impact excitation cross section ($\sigma$, $\pi{a_0^2}$) is related to dimensionless parameter collision strength ($\Omega$) as
\begin{equation}
\Omega_{ij}(E) = {k^2_i}\omega_i\sigma_{ij}(E)
\end{equation}
where ${k^2_i}$ is the incident energy of the electron and $\omega_i$ is the statistical weight of the initial state. The only transitions for which $\sigma$ have been measured, at energies up to 1.5~Ryd,  are 1s$^2$~$^1$S -- 3s3p~$^{1,3}$P$^o$ \citep{wall, rise}, and there is no discrepancy with theoretical results -- see figs. 2 and 3 of the former and fig. 4 of the latter, and also fig. 1 of \cite*{kai}, who adopted the earlier version of {\sc darc}.

Since very little data for $\Omega$ are available in the literature for transitions in Si~III, in Table~3 we list our results for all resonance transitions (i.e. from the ground to higher excited levels), at energies above thresholds but over a wide range of 4--30~Ryd. This should be useful for comparisons in future and for assessing the accuracy of our results. Regarding present comparisons, \cite{dk1} have listed their results for only a few transitions, and in Table~D we compare these with our calculations. The $\Omega$ of \cite{dk1} are nearly constant over a very wide energy range of 0.7 to 10~Ryd, but as expected $\Omega$ does vary with energy, irrespective of the type of transition, as is clear from our results listed  in Tables 3 and D. More surprisingly, for half of these (limited) transitions their results differ with ours by over an order of magnitude -- see for example, 1--9/10/13/14. In some instances their $\Omega$ values are higher and lower for others. Although their calculations (in comparison) are not very accurate, mainly because they included a limited range of partial waves with angular momentum $L \le$ 12, such large differences are not understandable. In the absence of any other results being available for comparisons, we have performed another calculation with {\sc fac} by including the same 141 levels as with {\sc darc}.

{\sc fac} is also a relativistic code, as stated earlier. It calculates collisional data with the {\em distorted-wave} (DW) method, and as has been demonstrated in several of our earlier papers the results for $\Omega$ are often comparable with those with {\sc darc} for most of the transitions, particularly at energies above thresholds. Therefore, in Table~D we have also listed DW $\Omega$, but at a single energy of $\sim$10~Ryd.  Discrepancies between the {\sc fac} and {\sc darc} $\Omega$ are up to 50\% (except for 1--14: 1s$^2$~$^1$S$_0$ -- 3p$^2$~$^1$S$_0$ for which differences are of a factor of two) for several transitions, but the agreement between the two independent calculations is much closer than with the earlier $R$-matrix results of \cite{dk1}. Therefore, the listed $\Omega$ of \cite{dk1} do not appear to be accurate. However, since it is the effective collision strengths (see next section) which are applied in the modelling of plasmas, it will be more useful to compare the $\Upsilon$ results to draw any meaningful conclusion.

\begin{figure*}
\includegraphics[angle=90,width=0.9\textwidth]{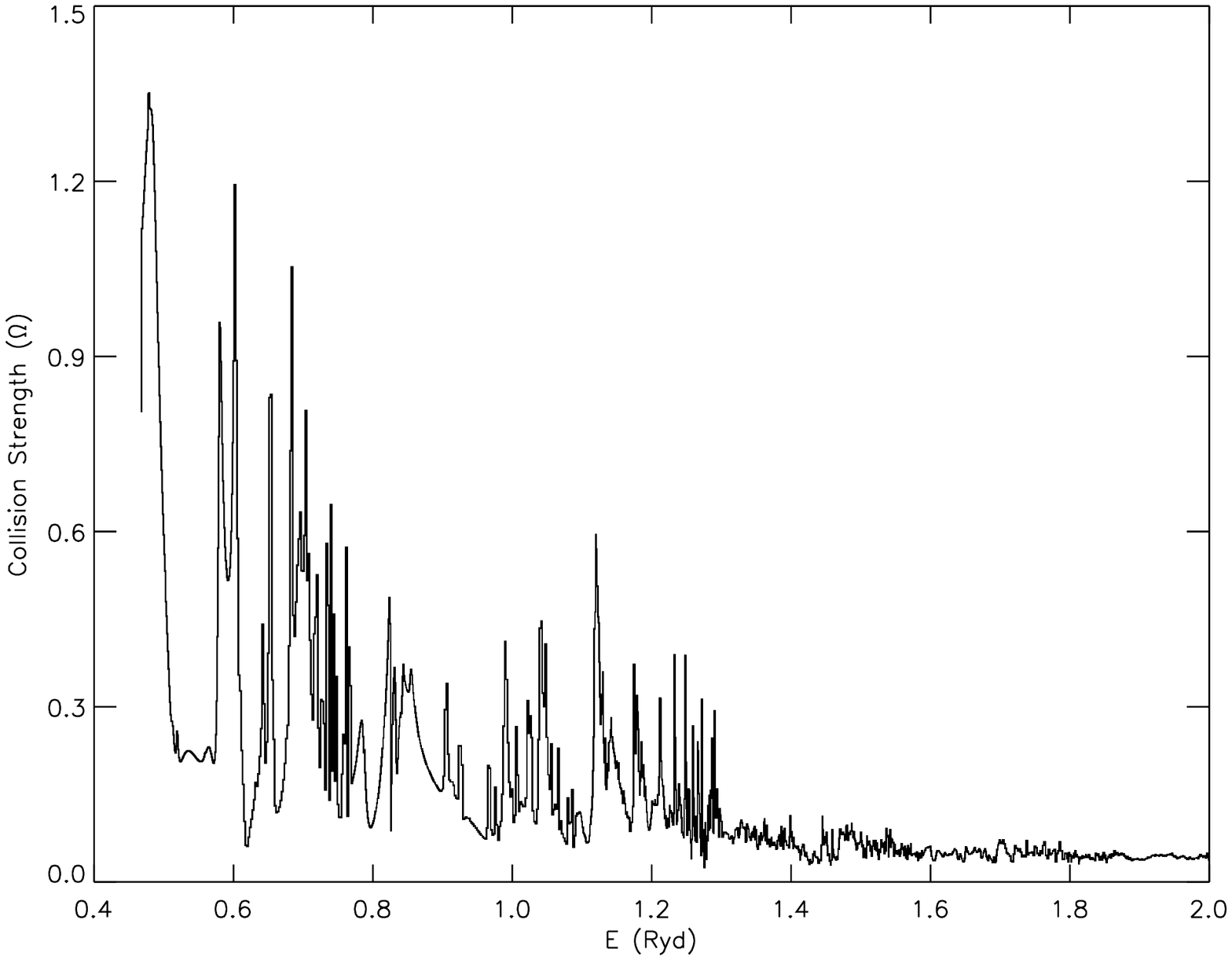}
 \vspace{-1.5cm}
 \caption{Collision strengths for the  1--2 (3s$^2$~$^1$S$_0$ -- 3s3p~$^3$P$^o_0$) transition of Si III.}
 \end{figure*}

\setcounter{figure}{1}
 \begin{figure*}
\includegraphics[angle=90,width=0.9\textwidth]{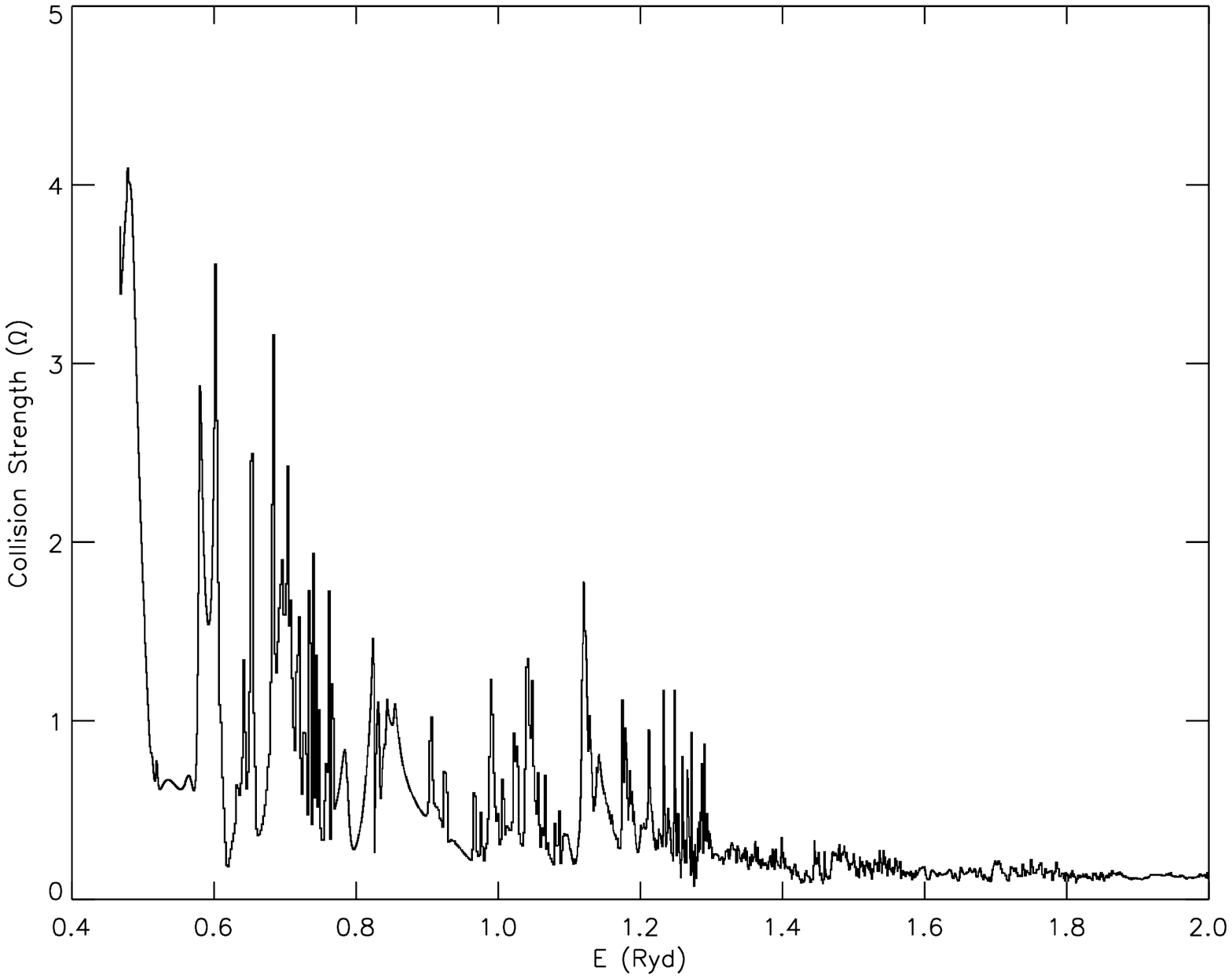}
 \vspace{-1.5cm}
\caption{Collision strengths for the 1--3 (3s$^2$~$^1$S$_0$ -- 3s3p~$^3$P$^o_1$) transition of Si III.}
 \end{figure*}

\setcounter{figure}{2}
 \begin{figure*}
\includegraphics[angle=90,width=0.9\textwidth]{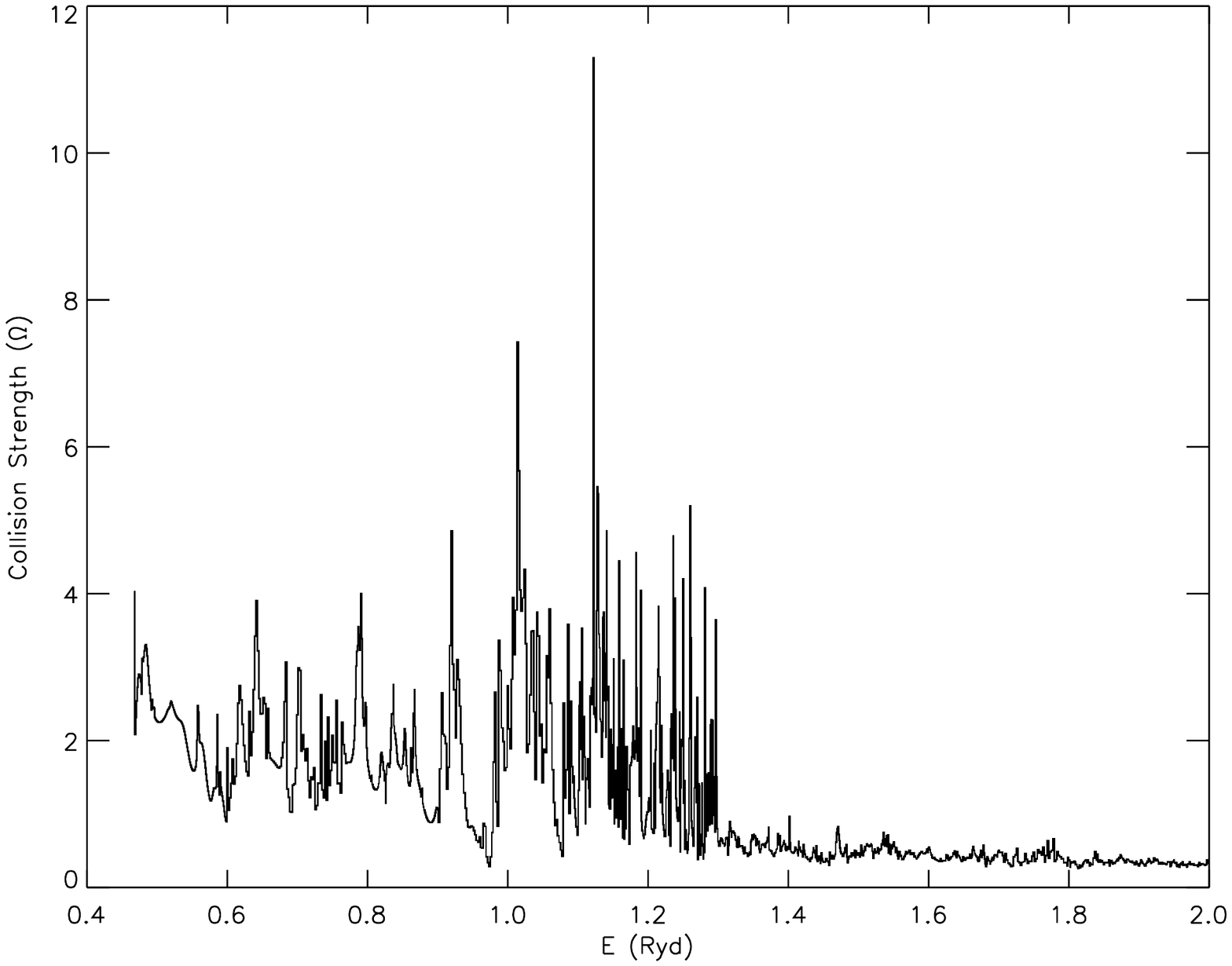}
 \vspace{-1.5cm}
 \caption{Collision strengths for the 2--3 (3s3p~$^3$P$^o_0$ -- 3s3p~$^3$P$^o_1$) transition of Si III.}
 \end{figure*}

\section{Effective collision strengths}

In the thresholds region values of $\Omega$ do not vary  smoothly because of the numerous closed-channel (Feshbach) resonances, and need to be resolved in a fine energy mesh.  However, resonances for transitions in Si~III are not as prominent as for other ions, and this can be judged from  figs. 2 and 3 of \cite{wall}, figs. 4 and 5 of \cite{rise}, fig. 1 of \cite{kai}, and  figs. 4--8 of \cite{dcg}. Nevertheless, in Figs. 1--3  we show resonances for three transitions, namely 1--2 (3s$^2$~$^1$S$_0$ -- 3s3p~$^3$P$^o_0$), 1--3 (3s$^2$~$^1$S$_0$ -- 3s3p~$^3$P$^o_1$) and 2--3 (3s3p~$^3$P$^o_0$ -- 3s3p~$^3$P$^o_1$). The 1--2 and 2--3 are forbidden whereas 1--3 is an inter-combination  (allowed) transition. Resonances in these figures are shown at energies below 2~Ryd, because $\Omega$ varies (almost) smoothly at higher energies. This may be the reason that \cite{dk1} provided average values of $\Omega$ in the 0.7 to 10~Ryd energy region -- see Table~D.  We have resolved resonances with an energy mesh of 0.001~Ryd in most of the thresholds region, and have calculated $\Omega$ at over 2600 points.

Because of the resonances, as shown in Figs. 1--3, values of $\Omega$ are averaged over a  {\em Maxwellian} distribution as follows: 

\begin{equation}
\Upsilon(T_e) = \int_{0}^{\infty} {\Omega}(E) \, {\rm exp}(-E_j/kT_e) \,d(E_j/{kT_e}),
\end{equation}
where $k$ is Boltzmann constant, T$_e$  the electron temperature in K, and E$_j$  the electron energy with respect to the final (excited) state. This value of $\Upsilon$ is
related to the excitation q(i,j) and de-excitation q(j,i) rates as follows:

\begin{equation}
q(i,j) = \frac{8.63 \times 10^{-6}}{{\omega_i}{T_e^{1/2}}} \Upsilon \, {\rm exp}(-E_{ij}/{kT_e}) \hspace*{1.0 cm}{\rm cm^3s^{-1}}
\end{equation}
and
\begin{equation}
q(j,i) = \frac{8.63 \times 10^{-6}}{{\omega_j}{T_e^{1/2}}} \Upsilon \hspace*{1.0 cm}{\rm cm^3 s^{-1}},
\end{equation}
where $\omega_i$ and $\omega_j$ are the statistical weights of the initial ($i$) and final ($j$) states, respectively, and E$_{ij}$ is the transition energy. Results for these rates are required in the modelling of plasmas.

 \setcounter{figure}{3} 
  \begin{figure*}
\includegraphics[angle=-90,width=0.9\textwidth]{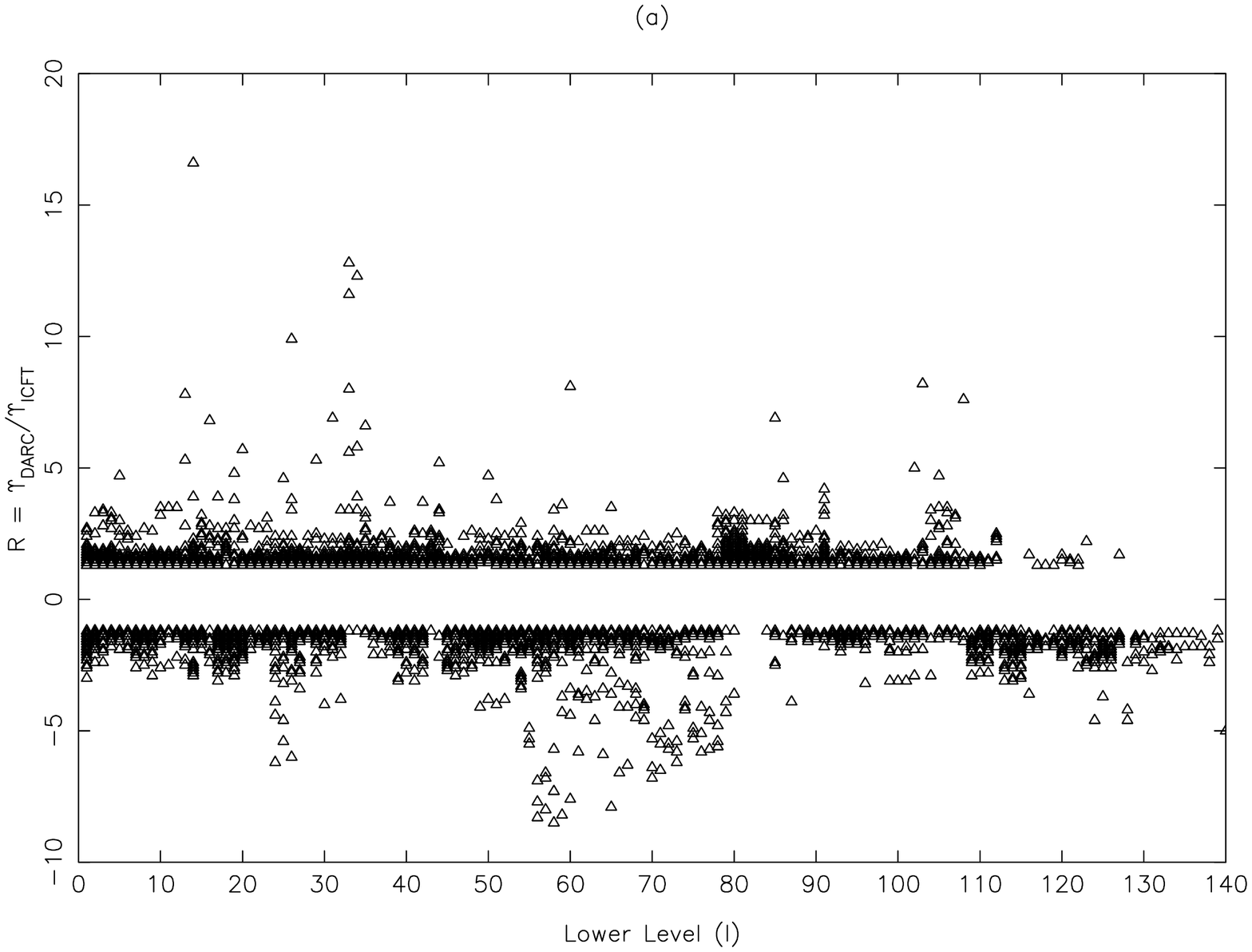}
 \vspace{-1.5cm}
\caption{Comparisons of  $\Upsilon$ between our results with {\sc darc} and those of \cite{icft1}  with ICFT for transitions of Si~III at (a) T$_e$ = 1.8$\times$10$^3$, (b) T$_e$ = 4.5$\times$10$^4$ and (c) T$_e$ = 1.8$\times$10$^6$~K. Negative R values indicate that $\Upsilon_{\rm DARC}$ $<$ $\Upsilon_{\rm ICFT}$. Only those transitions are shown which differ by over 20\%.}
 \end{figure*}

\setcounter{figure}{3} 
 \begin{figure*}
\includegraphics[angle=-90,width=0.9\textwidth]{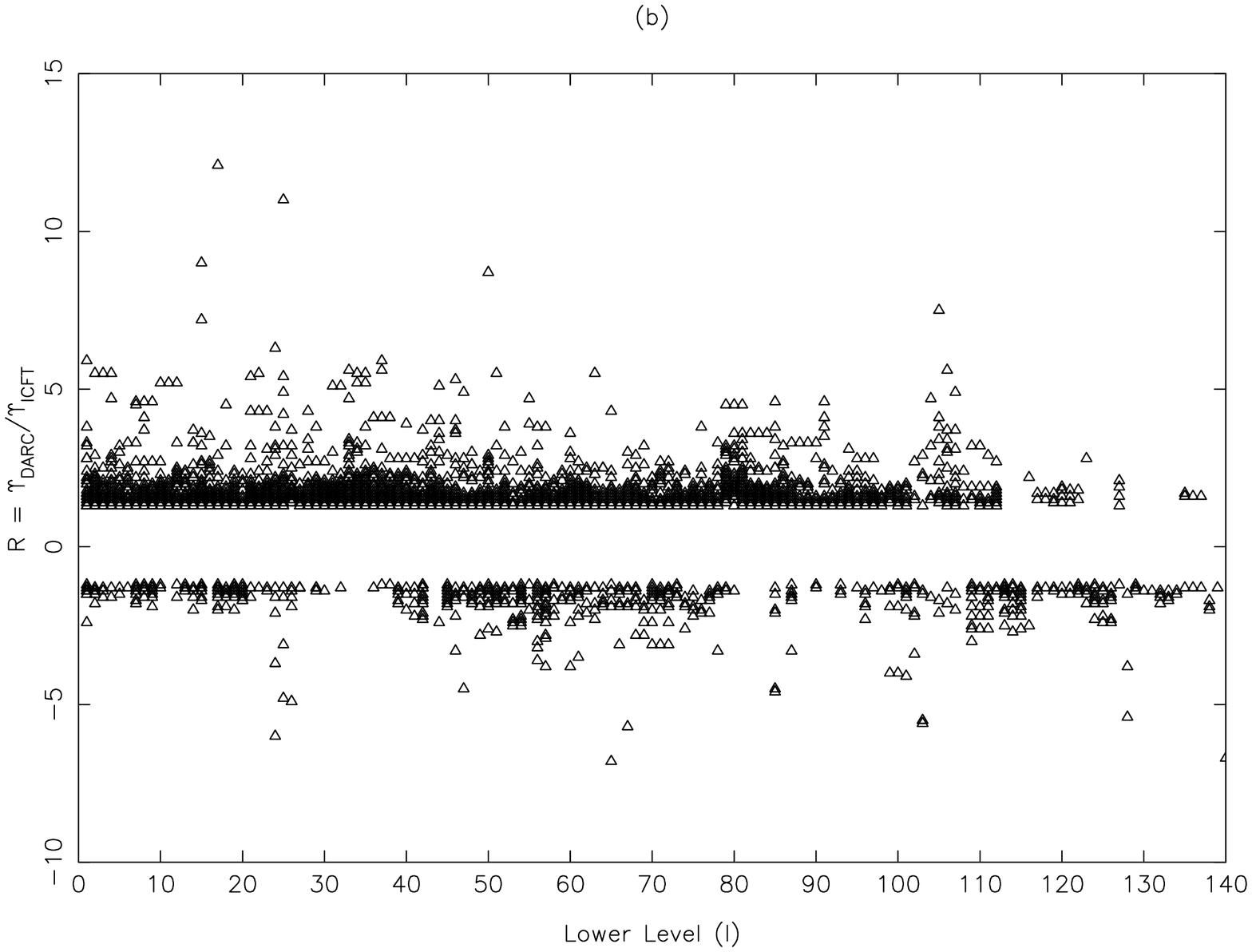}
 \vspace{-1.5cm}
 \caption{b. continued}
 \end{figure*}

\setcounter{figure}{3} 
 \begin{figure*}
\includegraphics[angle=-90,width=0.9\textwidth]{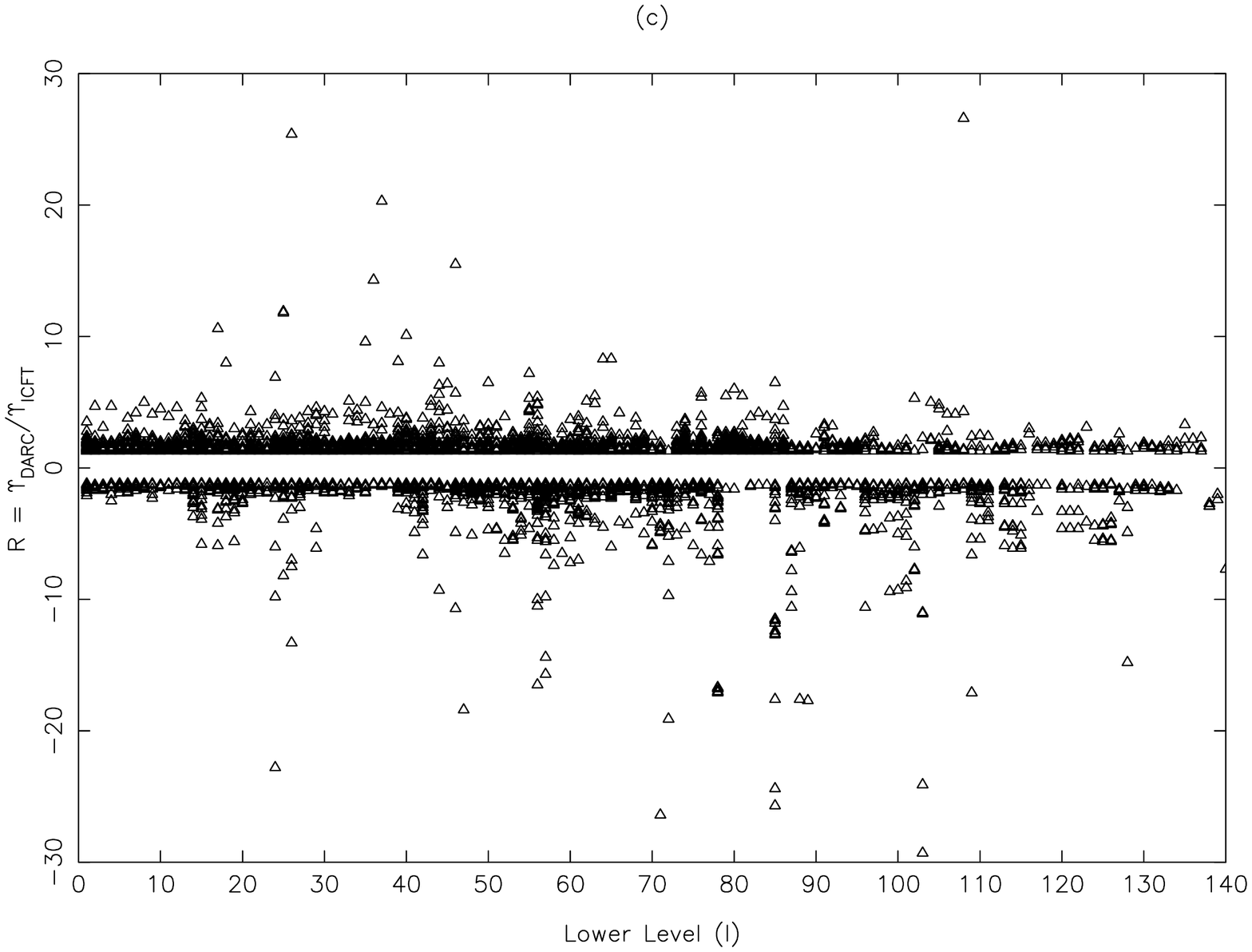}
 \vspace{-1.5cm}
 \caption{c. continued}
 \end{figure*}

 \setcounter{figure}{4}
\begin{figure*}
\includegraphics[angle=-90,width=0.9\textwidth]{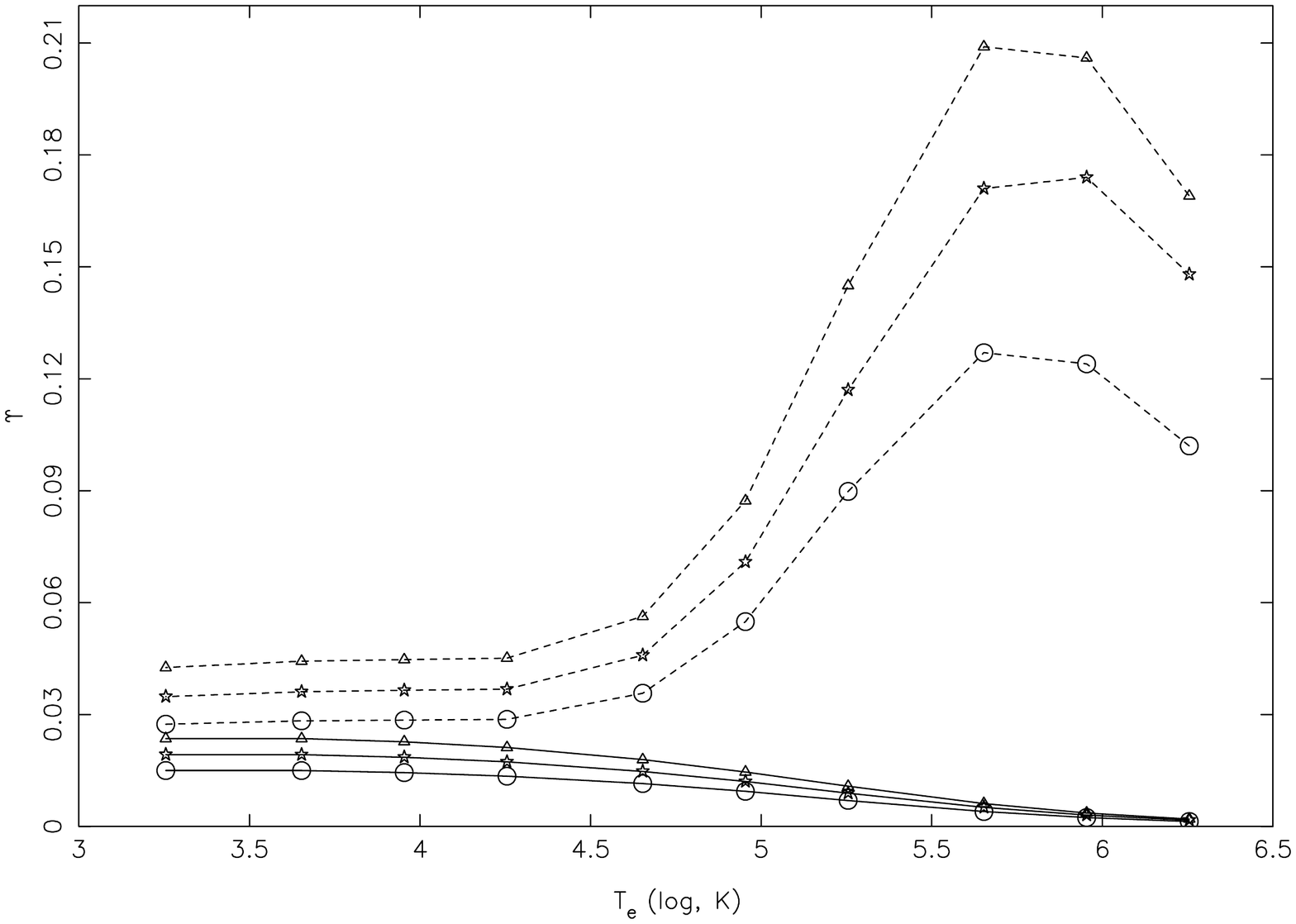}
 \vspace{-1.5cm}
 \caption{Comparison of effective collision strengths for the  70--141 (circles: 3p4f~$^3$G$_3$ -- 3d4d~$^1$S$_0$), 71--141 (stars: 3p4f~$^3$G$_4$ -- 3d4d~$^1$S$_0$) and 72--141 (triangles: 3p4f~$^3$G$_5$ -- 3d4d~$^1$S$_0$) forbidden transitions of Si~III. Continuous curves: present results with {\sc darc}, broken curves: ICFT results of  \cite{icft1}.}
 \end{figure*}
 
\setcounter{figure}{5} 
 \begin{figure*}
\includegraphics[angle=-90,width=0.9\textwidth]{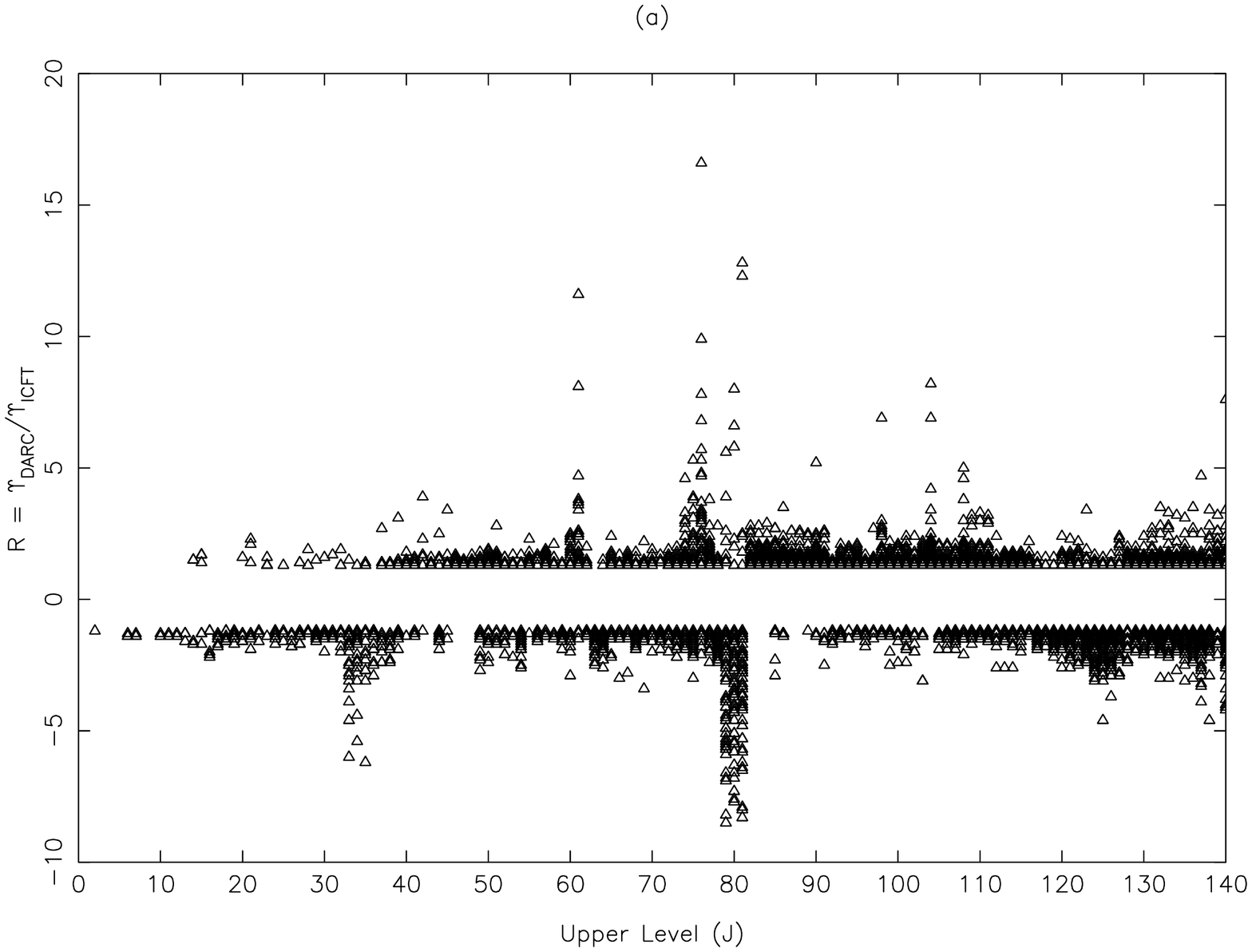}
 \vspace{-1.5cm}
\caption{Comparisons of  $\Upsilon$ between our results with {\sc darc} and those of \cite{icft1} with ICFT for transitions of Si~III at (a) T$_e$ = 1.8$\times$10$^3$, (b) T$_e$ = 4.5$\times$10$^4$ and (c) T$_e$ = 1.8$\times$10$^6$~K. Negative R values indicate that $\Upsilon_{\rm DARC}$ $<$ $\Upsilon_{\rm ICFT}$. Only those transitions are shown which differ by over 20\%.}
 \end{figure*}

\setcounter{figure}{5} 
 \begin{figure*}
\includegraphics[angle=-90,width=0.9\textwidth]{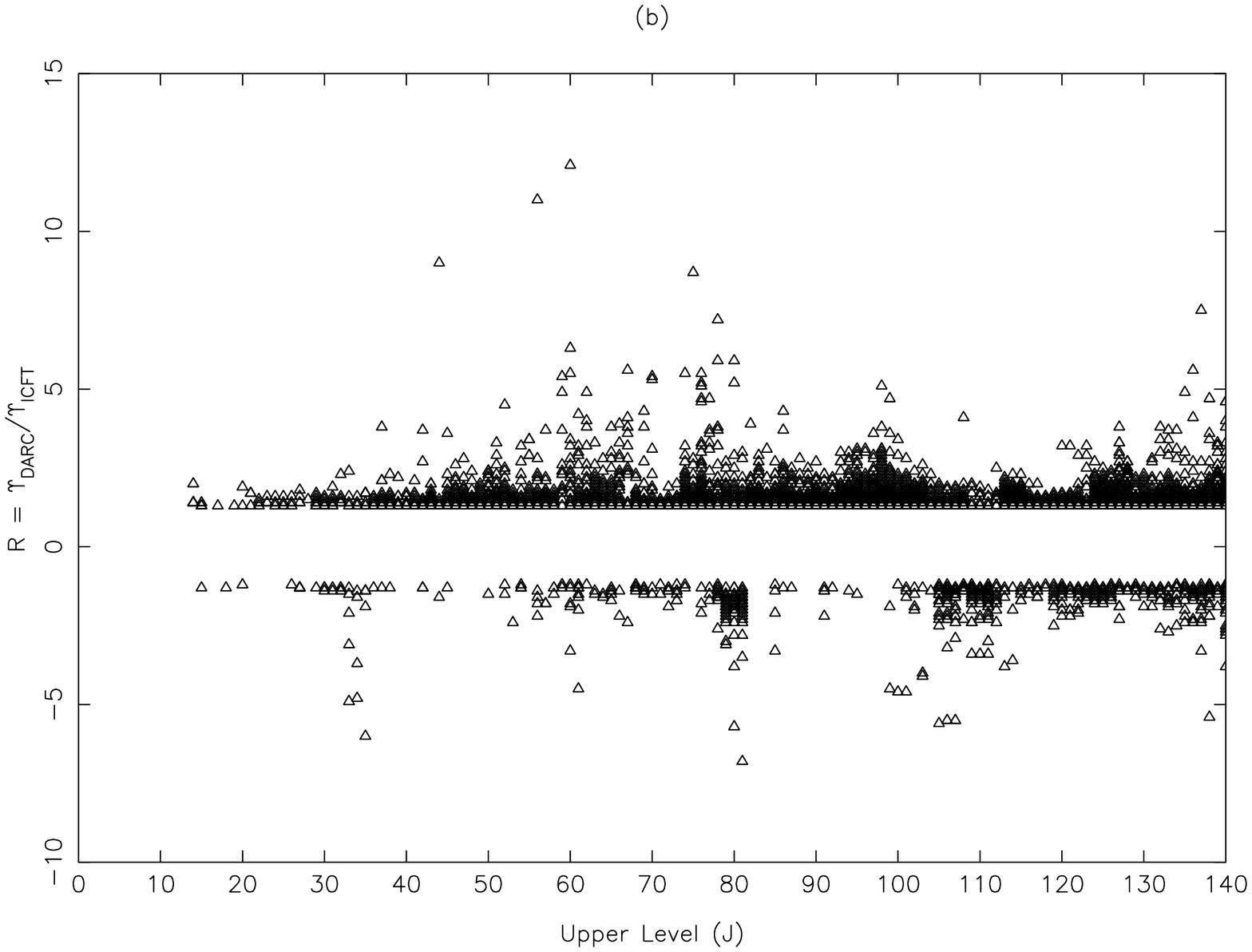}
 \vspace{-1.5cm}
 \caption{b. continued}
 \end{figure*}

\setcounter{figure}{5} 
 \begin{figure*}
\includegraphics[angle=-90,width=0.9\textwidth]{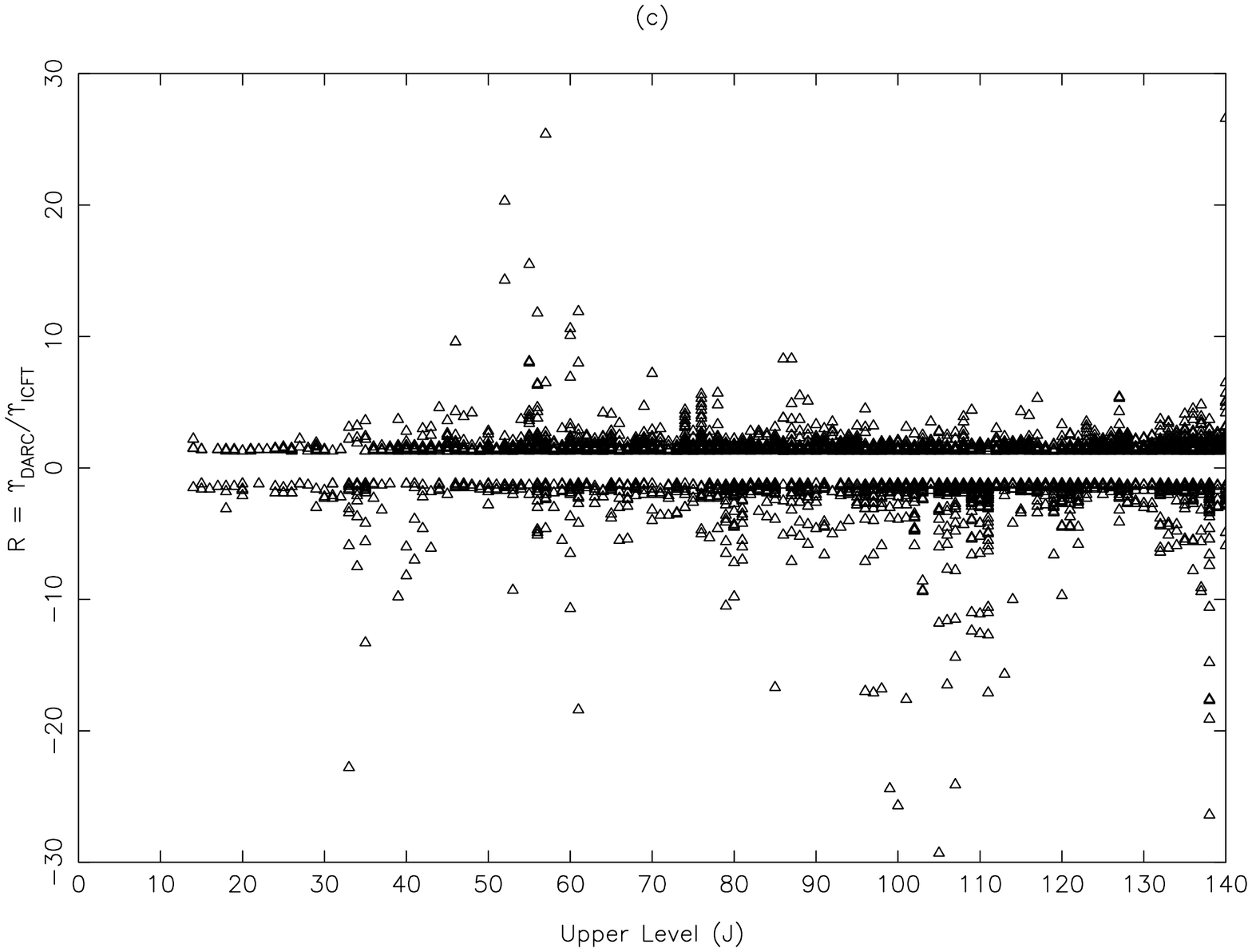}
 \vspace{-1.5cm}
 \caption{c. continued}  
 \end{figure*}
 
 Our calculated values of  $\Upsilon$ are listed in Table~4 at temperatures up to  10$^{5.9}$~K, well above  the T$_e$ of maximum abundance in ionisation equilibrium for Si~III, i.e. 10$^{4.7}$~K  \citep{pb}. However, for brevity only transitions from the lowest 29 to higher excited levels are listed in Table~4, but full table is  available online in the electronic version (see Appendix A).  As  discussed in section~1, the most recent, extensive and benchmarked \citep{bench} data for $\Upsilon$ are those of \cite{icft1}.  Therefore, we will undertake a detailed  comparison  with their results, but before that in Table~E  we make a short comparison with the other available data of \cite{dk1} and \cite{icft5} for the same transitions listed in Table~D, at the most relevant T$_e$ = 45~000~K. The only transition for which the $\Upsilon$ of \cite{dk1} differs substantially (by a factor of four) with our result is 1--14 (3s$^2$~$^1$S$_0$ -- 3p$^2$~$^1$S$_0$), and this is a direct consequence of their corresponding lower values of $\Omega$ as seen in Table~D. For other transitions, differences in $\Upsilon$ values are not as noticeable as for $\Omega$s, because T$_e$ = 45~000~K is equivalent to only 0.285~Ryd, whereas the comparisons of $\Omega$ shown in Table~D are at much higher energies.

The other results of $\Upsilon$ listed in Table~E are of \cite{icft5} and \cite{icft1}, i.e. ICFT1 and ICFT2, respectively. As stated earlier in section~1, the $\Upsilon$ results of \cite{icft5} for Mg-like ions were in error (see column ICFT1a in Table~E), but were subsequently corrected and  stored in the OPEN-ADAS database at {\tt http://www.open.adas.ac.uk} -- see column under ICFT1b. The ICFT1a and ICFT1b \,$\Upsilon$ differ by up to a factor of 35 for some transitions, such as 1--15 (3s$^2$~$^1$S$_0$ -- 3s4s~$^1$S$_0$). However, for the transitions listed in Table~E there are no great discrepancies between our DARC and earlier ICFT results, although the ICFT2 \,$\Upsilon$ of \cite{icft1} for the 1--14 (3s$^2$~$^1$S$_0$ -- 3p$^2$~$^1$S$_0$) is lower by a factor of two. We discuss comparisons with their results in detail below for a larger number of transitions and over a wider range of temperatures.

In Fig. 4 (a, b and c) we compare our $\Upsilon$ with the ICFT results of  \cite{icft1}. These are shown in the form of the ratio R = $\Upsilon_{DARC}$/$\Upsilon_{ICFT}$, with negative values of R representing $\Upsilon_{ICFT}$/$ \Upsilon_{DARC}$, i.e.  $\Upsilon_{ICFT}  >  \Upsilon_{DARC}$. These comparisons of $\Upsilon$ are for all 9870 transitions among the 141 levels, listed in Table~1. Fern{\'a}ndez-Menchero et al. \cite{icft1} have calculated results among 283 levels and therefore we have carefully isolated the common levels/transitions from their work. The comparisons shown in Fig.~4 are   at  three temperatures of 10$^{3.255}$, 10$^{4.653}$ and 10$^{6.255}$~K. The first and the third are the lowest and the highest common temperatures between the two calculations, whereas the second is the most relevant for applications to astrophysical plasmas.

At T$_e$ = 10$^{3.255}$~K, about half the transitions differ by over 20\%, and among these for about half $\Upsilon_{DARC}  >  \Upsilon_{ICFT}$ and for the other $\Upsilon_{ICFT}  >  \Upsilon_{DARC}$. Since 10$^{3.255}$~K is a very low temperature ($\sim$ 0.011~Ryd), differences as seen in Fig. 4a are common between any two independent calculations, because the position of resonances can significantly affect the magnitudes of $\Upsilon$. Similar discrepancies (for 55\% of transitions) between the two sets of $\Upsilon$ are seen in Fig. 4b  at T$_e$ = 10$^{4.653}$~K, although a much better agreement is expected.  

Unfortunately, discrepancies for about 50\% of transitions remain at T$_e$ = 10$^{6.255}$~K, equivalent to $\sim$11.4~Ryd, a temperature well beyond the highest threshold considered in any of the two calculations, and at which the contributions of resonances, if any, are not expected to be significant. In fact, Fern{\'a}ndez-Menchero et al. \cite{icft1} have also concluded that the effect of resonances attached to higher excited levels is not significant. Therefore, the larger calculations performed by \cite{icft1}  cannot be the reason for such large discrepancies. More importantly, the magnitude of discrepancies is much larger at this temperature, and for a majority of transitions $\Upsilon_{ICFT}  >  \Upsilon_{DARC}$. Additionally, there are 13 transitions which are out of scale in Fig.~4c, and these are: 24--33/34/35, 25--33/34, 26--33, 102--104/102 and 103--106 (all allowed), and 70/71/72 -- 141 and 102--111, forbidden. Invariably for all these transitions, $\Upsilon_{ICFT}  >  \Upsilon_{DARC}$ by up to (almost) two orders of magnitude.

Some differences for a few, particularly allowed and inter-combination transitions, are understandable. For example, for the 24 -- 33, 34, 35 (3s4d~$^3$D$_1$ -- 3p3d~$^3$P$^o_{2,1,0}$)  transitions, our A-values are 5.81$\times$10$^2$ (f = 5.56$\times$10$^{-6}$), 7.70$\times$10$^3$ (f = 4.35$\times$10$^{-5}$) and 2.33$\times$10$^4$ (f = 4.35$\times$10$^{-5}$)~s$^{-1}$, respectively, i.e. all such transitions are very {\em weak}. Subsequently, as expected, our values of $\Omega$ for such transitions have fully converged within the adopted $J$ range of $\le$ 40.5, and both $\Omega$ and $\Upsilon$ decrease with increasing energy/temperature. However, the corresponding A-values of \cite{icft1} from the {\sc as} calculations are 1.06$\times$10$^5$, 2.30$\times$10$^6$ and 8.69$\times$10$^6$~s$^{-1}$, respectively, i.e. higher by up to three orders of magnitude. Consequently, these transitions in their calculations are much stronger and may have higher magnitude of $\Omega$ and $\Upsilon$. Nevertheless, these (and other) transitions remain weak and may not necessarily follow the f-values, because weaker transitions often behave as forbidden.

Therefore, the discrepancies become clearer when we have a closer look at some of the forbidden transitions, such as 70/71/72 -- 141 (3p4f~$^3$G$_{3,4,5}$ --  3d4d~$^1$S$_0$), which correspond to 90/92/93 -- 216 in the calculations of \cite{icft1}, and for resonances cover a narrow energy region of 0.2~Ryd between their 217 and 283 threshold levels, which cannot be a major source of enhancement  in $\Upsilon$ values at higher temperatures. However, for these (and many other) transitions their $\Upsilon$ increase with increasing T$_e$ (up to about 10$^6$~K and then decrease), whereas our results continuously decrease, as expected. The differences in $\Upsilon$ results can be better appreciated from Fig. 5, in which the behaviour of the ICFT \,$\Upsilon$ is not correct, and this is because of the extrapolation of their $\Omega$ over a very large energy range, as stated earlier in the paper.

Finally, we make one more comparison in Fig.~6 (a, b and c) at the same three temperatures as in Fig.~4, but this time replacing the lower levels (I) with upper ones (J), because this provides a clearer picture of the similarities or differences among transitions  up to level(s) J. Indeed this figure is more revealing than Fig.~4, because only for transitions among the lowest 29 levels there are no large discrepancies at any temperature between our DARC and the ICFT calculations of \cite{icft1}. However, discrepancies increase with increasing T$_e$ and become worse as J increases. At the lowest T$_e$, there is a reasonable agreement between the two calculations for transitions with J $\le$ $\sim$70, which decreases to $\sim$ 50 and 30 as T$_e$ increases, as seen in Fig.~6 (b and c). Therefore, as discussed earlier the maximum problem is at temperatures towards the higher end, and  for a majority of transitions $\Upsilon_{ICFT}  >  \Upsilon_{DARC}$. This conclusion is consistent with that observed earlier for Be-like ions \citep{alx1, ciii, niv}.

\section{Conclusions}

In this paper we have reported energies and lifetimes, calculated with the {\sc grasp} code, for the 141 levels of the  3$\ell3\ell'$ and 3$\ell$4$\ell$ configurations of Si~III. Experimental energies are available for only the lowest 58, but there is no major discrepancy with theoretical results, either in magnitude or orderings.   Furthermore, increasing the CI does not make the energy levels more accurate, but levels arising from higher configurations intermix with those considered here.  Radiative rates, particularly for E1 transitions, also show a good agreement among various calculations, but only for comparatively strong transitions (f $>$ 0.1). For weaker transitions, differences among different calculations are up to a factor of three, mainly because of differing methodologies and CI. For such transitions, it is difficult to assess the accuracy with confidence. However, our lifetimes show a reasonably satisfactory agreement with other available theoretical results as well as the measurements, and therefore provide some confidence in the calculations. 

Considering the same 141 levels, as for radiative rates, we have also calculated collision strengths  with the fully relativistic {\sc darc} code. The calculated results are listed for all resonance transitions over a wide  energy range,  up to 30~Ryd. These results should be useful for future comparisons, because no similar data are currently available in the literature. Resonances in a narrow energy mesh have also been resolved in the thresholds region to determine $\Upsilon$ values, which are required for the diagnostics and modelling of plasmas. Results are listed over a large range of  temperatures up to 10$^{5.9}$~K, for all 9870 transitions among the 141 levels. Similar results for a larger range of transitions, among 283 levels,  and with the $R$-matrix code are available \citep{icft1}. However, the earlier calculations are primarily in the $LS$ coupling and results for fine-structure transitions have been determined through the ICFT approach. More importantly, these earlier calculations for $\Omega$ have been performed over a limited energy range (below 7.4~Ryd), and have been extrapolated over a very wide energy range to calculate subsequent results for $\Upsilon$ up to T$_e$ = 1.8$\times$10$^7$~K, which amounts to 114~Ryd.   As a result of this discrepancies between the two sets of data are of over 20\% for about half the transitions, and at all temperatures. In general, discrepancies increase with increasing temperature and are more prominent for transitions belonging to higher levels (J $>$ 29). In a majority of cases the earlier ICFT results of $\Upsilon$ are higher, by up to (almost) two orders of magnitude. 

Since \cite{icft1} have performed larger calculations, some differences with our work are expected and understandable, because for some allowed transitions the f-values differ, and for forbidden ones a larger range of resonances has been included. However,  resonances  for transitions in Si~III are not very important, as seen here in Figs. 1--3 and also confirmed earlier by \cite{icft1} and other workers, and therefore their contribution alone (if any) cannot explain the large discrepancies noted (particularly) for the forbidden transitions, as shown in Fig.~5.  Furthermore, the span of energy range for the additional levels included by them is very small, i.e. only 0.2~Ryd, which may affect the values of $\Upsilon$ at low temperatures, but not the higher ones discussed in the paper. The conclusion that their methodology leads to significant overestimation in the determination of $\Upsilon$ values is consistent with that derived earlier for Be-like ions. Therefore, we believe, the presently reported results of $\Upsilon$ for transitions in Si~III are more accurate than currently available in the literature, and hence should be adopted in the modelling of plasmas. Furthermore, since \cite{icft1} have adopted the same methodology for all Mg-like ions, up to Z = 36, it is advisable to perform revised calculations for all ions so that the data can be confidently applied for modelling and/or diagnostics of plasmas. 

Our presented results for $\Upsilon$ are assessed to be more accurate than the existing ones, but scope remains for improvement. This is mainly because levels arising from the $n \ge$ 5 configurations highly mix with those of $n \le$ 4 considered in the present work, but have been omitted due to practical (computational) reason. Their inclusion may  improve the accuracy of the wavefunctions, and will certainly lead to the more accurate determination of  $\Upsilon$ results. However, the presently listed results for transitions belonging to the lowest 29 levels of Si~III can be confidently applied, because these are likely to be unaffected by the inclusion of levels from higher configurations.     \\ \\
 \newpage
 
 {\bf Appendix A. Supplementary data}\\
Supplementary material related to this article can be found online at http://dx.doi.org/00.0000/j.adt.2017.00.001. Owing to space limitations, only parts of Tables 2 and 4 are presented here, but full tables are being made available as supplemental material in conjunction with the electronic publication of this work.

\clearpage

\renewcommand{\baselinestretch}{1.5}

\footnotesize  

\begin{longtable}{rrrrrrrrrrrrr}
\caption{\label{table_A}
Comparison of f-values for E1 transitions among the lowest 22 levels of Si~III.  $a{\pm}b \equiv a{\times}$10$^{{\pm}b}$.   See Table 1 for level indices.
}
Transition & & GRASP1 & & GRASP2 & & FAC1 & FAC2 & MCHF \\

\hline
I & J & f & R & f & R & f & f & f \\
\hline
\endfirsthead
\caption[]{(continued)}
Transition & & GRASP1 & & GRASP2 & & FAC1 & FAC2 & MCHF \\

\hline
I & J & f & R & f & R & f & f & f \\
\hline
\endhead
    1   &   3   &     1.820$-$5   &  8.9$-$1  &   1.929$-$5  &   9.1$-$1  &  1.645$-$5  &  1.854$-$5  &  2.816$-$5  \\
     1   &   5   &     1.679$-$0   &  9.7$-$1  &   1.691$-$0  &   9.7$-$1  &  1.683$-$0  &  1.691$-$0  &  1.606$-$0  \\
     1   &  18   &     4.196$-$5   &  4.7$-$1  &   3.254$-$5  &   7.7$-$1  &  2.372$-$5  &  2.462$-$5  &  7.111$-$5  \\
     1   &  20   &     3.579$-$2   &  2.7$-$1  &   1.775$-$2  &   6.5$-$1  &  2.062$-$2  &  1.659$-$2  &  2.241$-$2  \\
     2   &   8   &     5.551$-$1   &  1.0$-$0  &   5.591$-$1  &   1.0$-$0  &  5.538$-$1  &  5.572$-$1  &  5.328$-$1  \\
     2   &  12   &     8.804$-$1   &  9.8$-$1  &   8.836$-$1  &   9.8$-$1  &  8.788$-$1  &  8.817$-$1  &  8.592$-$1  \\
     2   &  13   &     1.182$-$1   &  9.0$-$1  &   1.186$-$1  &   9.2$-$1  &  1.281$-$1  &  1.267$-$1  &  1.189$-$1  \\
     3   &   6   &     8.544$-$5   &  9.9$-$1  &   8.798$-$5  &   9.9$-$1  &  7.680$-$5  &  8.211$-$5  &  1.250$-$4  \\
     3   &   7   &     1.841$-$1   &  1.0$-$0  &   1.854$-$1  &   1.0$-$0  &  1.836$-$1  &  1.847$-$1  &  1.766$-$1  \\
     3   &   8   &     1.386$-$1   &  1.0$-$0  &   1.396$-$1  &   1.0$-$0  &  1.383$-$1  &  1.390$-$1  &  1.329$-$1  \\
     3   &   9   &     2.315$-$1   &  1.0$-$0  &   2.331$-$1  &   1.0$-$0  &  2.309$-$1  &  2.324$-$1  &  2.223$-$1  \\
     3   &  11   &     6.599$-$1   &  9.8$-$1  &   6.623$-$1  &   9.8$-$1  &  6.587$-$1  &  6.614$-$1  &  6.440$-$1  \\
     3   &  12   &     2.201$-$1   &  9.8$-$1  &   2.209$-$1  &   9.8$-$1  &  2.197$-$1  &  2.206$-$1  &  2.148$-$1  \\
     3   &  13   &     1.188$-$1   &  9.0$-$1  &   1.191$-$1  &   9.2$-$1  &  1.287$-$1  &  1.272$-$1  &  1.194$-$1  \\
     3   &  14   &     7.773$-$6   &  9.8$-$1  &   8.372$-$6  &   9.5$-$1  &             &             &  1.497$-$5  \\
     3   &  15   &     3.526$-$6   &  7.4$-$1  &   3.325$-$6  &   7.7$-$1  &  2.335$-$6  &  2.462$-$6  &  1.397$-$6  \\
     3   &  16   &     3.043$-$5   &  8.6$-$1  &   2.964$-$5  &   9.0$-$1  &             &             &  3.972$-$5  \\
     4   &   6   &     8.856$-$5   &  1.0$-$0  &   9.149$-$5  &   1.0$-$0  &  7.592$-$5  &  7.947$-$5  &  1.336$-$4  \\
     4   &   8   &     1.375$-$1   &  1.0$-$0  &   1.385$-$1  &   1.0$-$0  &  1.372$-$1  &  1.380$-$1  &  1.319$-$1  \\
     4   &   9   &     4.147$-$1   &  1.0$-$0  &   4.177$-$1  &   1.0$-$0  &  4.138$-$1  &  4.162$-$1  &  3.979$-$1  \\
     4   &  10   &     7.385$-$1   &  9.8$-$1  &   7.413$-$1  &   9.8$-$1  &  7.373$-$1  &  7.402$-$1  &  7.209$-$1  \\
     4   &  11   &     1.320$-$1   &  9.8$-$1  &   1.325$-$1  &   9.8$-$1  &  1.318$-$1  &  1.323$-$1  &  1.289$-$1  \\
     4   &  12   &     8.803$-$3   &  9.8$-$1  &   8.837$-$3  &   9.8$-$1  &  8.788$-$3  &  8.828$-$3  &  8.596$-$3  \\
     4   &  13   &     1.199$-$1   &  9.0$-$1  &   1.203$-$1  &   9.3$-$1  &  1.298$-$1  &  1.285$-$1  &  1.206$-$1  \\
     4   &  16   &     2.967$-$6   &  9.8$-$1  &   3.168$-$6  &   9.6$-$1  &             &             &  4.452$-$6  \\
     5   &   6   &     4.502$-$2   &  8.1$-$1  &   4.487$-$2  &   8.5$-$1  &  4.022$-$2  &  4.074$-$2  &  4.785$-$2  \\
     5   &   7   &     1.229$-$5   &  1.2$-$0  &   1.241$-$5  &   1.2$-$0  &             &             &  1.476$-$5  \\
     5   &   8   &     1.374$-$6   &  1.0$-$0  &   1.470$-$6  &   1.0$-$0  &             &             &  2.153$-$6  \\
     5   &   9   &     1.260$-$5   &  1.0$-$0  &   1.242$-$5  &   1.0$-$0  &             &             &  1.845$-$5  \\
     5   &  11   &     1.379$-$5   &  1.0$-$0  &   1.433$-$5  &   1.0$-$0  &             &             &  1.907$-$5  \\
     5   &  12   &     4.068$-$6   &  9.4$-$1  &   4.315$-$6  &   9.3$-$1  &             &             &  6.200$-$6  \\
     5   &  13   &     4.739$-$6   &  1.1$-$0  &   5.113$-$6  &   1.0$-$0  &             &             &  7.021$-$6  \\
     5   &  14   &     3.178$-$1   &  1.1$-$0  &   3.118$-$1  &   1.1$-$0  &  2.982$-$1  &  2.890$-$1  &  2.238$-$1  \\
     5   &  15   &     5.822$-$3   &  2.6$-$1  &   7.979$-$3  &   4.3$-$1  &  2.852$-$2  &  3.379$-$2  &  7.637$-$2  \\
     5   &  16   &     1.715$-$0   &  8.5$-$1  &   1.689$-$0  &   8.9$-$1  &  1.688$-$0  &  1.687$-$0  &  1.644$-$0  \\
     6   &  18   &     5.406$-$5   &  9.2$-$1  &   6.939$-$5  &   9.4$-$1  &  4.095$-$5  &  5.591$-$5  &  1.534$-$4  \\
     6   &  19   &     7.683$-$8   &  1.5$-$0  &   5.446$-$8  &   1.5$-$0  &             &             &  0.409$-$9  \\
     6   &  20   &     9.952$-$2   &  9.3$-$1  &   9.763$-$2  &   9.5$-$1  &  1.065$-$1  &  1.033$-$1  &  8.917$-$2  \\
     6   &  21   &     1.496$-$4   &  9.9$-$1  &   1.434$-$4  &   9.9$-$1  &  1.440$-$4  &  1.372$-$4  &  1.737$-$4  \\
     6   &  22   &     3.152$-$5   &  8.1$-$1  &   3.627$-$5  &   8.5$-$1  &  3.376$-$5  &  3.900$-$5  &  5.625$-$5  \\
\hline               
\end{longtable}

\begin{flushleft}
GRASP1: present calculations from the {\sc grasp} code for 141 levels \\ 
GRASP2: present calculations from the {\sc grasp} code for 283 levels  \\ 
FAC1: present calculations from the {\sc fac} code for 283 levels  \\   
FAC2: present calculations from the {\sc fac} code for 1211 levels \\                                                                                          
R: ratio of velocity/length of f- values \\
MCHF: calculations of \cite{cff} with the {\sc mchf} code and available on the website: {\tt  http://nlte.nist.gov/MCHF/view.html}
\end{flushleft}


\begin{longtable}{rrrrrrr}
\caption{\label{table_B}
Comparison of oscillator strengths (f-values) for  some E2, M1 and M2 transitions of Si III.  $a{\pm}b \equiv a{\times}$10$^{{\pm}b}$.   See Table 1 for level indices.
}
I & J & Type & GRASP1 & GRASP2 & MCHF \\
\hline
\endfirsthead\\
\caption[]{(continued)}
I & J & Type & GRASP1 & GRASP2 & MCHF \\
\hline
\endhead
1  &  4  &  M2  &  3.091$-$11  &  3.129$-$11  &  3.358$-$11  \\
2  &  3  &  M1  &  9.968$-$09  &  9.987$-$09  &  1.042$-$08  \\
2  &  4  &  E2  &  1.331$-$13  &  1.355$-$13  &  1.464$-$13  \\
2  &  5  &  M1  &  6.335$-$11  &  6.447$-$11  &  7.896$-$11  \\
3  &  4  &  E2  &  3.021$-$14  &  3.075$-$14  &  3.329$-$14  \\
3  &  4  &  M1  &  8.478$-$09  &  8.495$-$09  &  8.881$-$09  \\
3  &  5  &  E2  &  2.867$-$12  &  2.985$-$12  &  3.510$-$12  \\
3  &  5  &  M1  &  1.612$-$11  &  1.690$-$11  &  2.002$-$11  \\
4  &  5  &  E2  &  1.255$-$12  &  1.270$-$12  &  1.407$-$12  \\
4  &  5  &  M1  &  1.565$-$11  &  1.591$-$11  &  1.949$-$11  \\
\hline   
\end{longtable}
\begin{flushleft}
GRASP1: present calculations from the {\sc grasp} code for 141 levels \\ 
GRASP2: present calculations from the {\sc grasp} code for 283 levels \\ 
MCHF: calculations of \cite{cff} with the {\sc mchf} code and available on the website: {\tt  http://nlte.nist.gov/MCHF/view.html}
\end{flushleft}

\clearpage


\begin{longtable}{rllllllllll}
\caption{\label{table_C}
Comparison of lifetimes ($\tau$, s) for the  lowest 29 levels of Si~III.  $a{\pm}b \equiv a{\times}$10$^{{\pm}b}$.   See Table 1 for level indices.
}
Index &
\multicolumn{1}{c}{Configuration} &
\multicolumn{1}{c}{Level} &
\multicolumn{1}{c}{GRASP1} &
\multicolumn{1}{c}{MCHF1} &
\multicolumn{1}{c}{MCHF2} &
\multicolumn{1}{c}{MBPT} &
\multicolumn{1}{c}{Exp.1} &
\multicolumn{1}{c}{Exp2.} \\
\hline
\endfirsthead\\
\caption[]{(continued)}
Index &
\multicolumn{1}{c}{Configuration} &
\multicolumn{1}{c}{Level} &
\multicolumn{1}{c}{GRASP1} &
\multicolumn{1}{c}{MCHF1} &
\multicolumn{1}{c}{MCHF2} &
\multicolumn{1}{c}{MBPT} &
\multicolumn{1}{c}{Exp.1} &
\multicolumn{1}{c}{Exp2.} \\
\hline
\endhead
     1   &  3s$^2$   &  $^1$S$_0  $  &               &              &              &             &                                                  \\
    2   &  3s3p     &   $^3$P$_0^o$  &               &              &              &             &                                                  \\
    3   &  3s3p     &   $^3$P$_1^o$  &   9.448$-$05  & 5.809$-$05   &  5.718$-$05  & 1.01$-$04   &      (5.99 $\pm$ 0.36)$-$05$^{\rm Expt.3}$    \\
    4   &  3s3p     &   $^3$P$_2^o$  &   9.003$+$01  & 7.846$+$01   &              &             &                                                  \\
    5   &  3s3p     &   $^1$P$_1^o$  &   3.763$-$10  & 4.050$-$10   &  4.077$-$10  & 4.29$-$10   &                                                  \\
    6   &  3p$^2$   &   $^1$D$_2  $  &   4.359$-$08  & 3.273$-$08   &  3.301$-$08  & 4.57$-$08   & (2.60 $\pm$ 0.15)$-$08 &  (2.6 $\pm$ 0.3)$-$08   \\
    7   &  3p$^2$   &   $^3$P$_0  $  &   4.547$-$10  & 4.779$-$10   &  4.792$-$10  & 4.65$-$10   &                                                  \\
    8   &  3p$^2$   &   $^3$P$_1  $  &   4.535$-$10  & 4.764$-$10   &  4.777$-$10  & 4.74$-$10   &                                                  \\
    9   &  3p$^2$   &   $^3$P$_2  $  &   4.514$-$10  & 4.741$-$10   &  4.753$-$10  & 4.21$-$10   &                                                  \\
   10   &  3s3d     &   $^3$D$_3  $  &   3.403$-$10  & 3.506$-$10   &  3.590$-$10  & 3.77$-$10   &                                                  \\
   11   &  3s3d     &   $^3$D$_2  $  &   3.387$-$10  & 3.524$-$10   &  3.608$-$10  & 3.74$-$10   &                                                  \\
   12   &  3s3d     &   $^3$D$_1  $  &   3.375$-$10  & 3.494$-$10   &  3.577$-$10  & 3.74$-$10   &                                                  \\
   13   &  3s4s     &   $^3$S$_1  $  &   4.195$-$10  & 4.441$-$10   &  4.132$-$10  &             &      (5.0$\pm$1.0)$-$10$^{\rm Expt.4}$               
                            \\
   14   &  3p$^2$   &   $^1$S$_0  $  &   3.272$-$10  & 4.068$-$10   &  4.983$-$10  & 4.76$-$10   & (5.8$\pm$0.4)$-$10$^{\rm Expt.5}$                    
                            \\
   15   &  3s4s     &   $^1$S$_0  $  &   1.514$-$08  & 1.112$-$09   &  1.127$-$09  &             &                                                  \\
   16   &  3s3d     &   $^1$D$_2  $  &   1.980$-$10  & 2.170$-$10   &  2.213$-$10  & 2.32$-$10   &                                                  \\
   17   &  3s4p     &   $^3$P$_0^o$  &   3.561$-$09  & 3.409$-$09   &  3.365$-$09  &             & (3.3 $\pm$ 0.3)$-$09   &  (4.1 $\pm$ 0.5)$-$09   \\  
   18   &  3s4p     &   $^3$P$_1^o$  &   3.545$-$09  & 3.390$-$09   &  3.347$-$09  &             & (3.6 $\pm$ 0.3)$-$09   &  (4.5 $\pm$ 0.5)$-$09   \\  
   19   &  3s4p     &   $^3$P$_2^o$  &   3.525$-$09  & 3.373$-$09   &  3.330$-$09  &             &                                                  \\
   20   &  3s4p     &   $^1$P$_1^o$  &   1.587$-$09  & 1.926$-$09   &  1.939$-$09  &             &                                                  \\
   21   &  3p3d     &   $^3$F$_2^o$  &   9.934$-$07  & 1.124$-$06   &  1.326$-$06  & 2.25$-$08   &                                                  \\
   22   &  3p3d     &   $^3$F$_3^o$  &   1.311$-$06  & 2.210$-$06   &  3.010$-$06  & 2.49$-$08   &                                                  \\
   23   &  3p3d     &   $^3$F$_4^o$  &   9.585$-$07  & 2.378$-$06   &  3.166$-$06  & 2.57$-$08   &                                                  \\
   24   &  3s4d     &   $^3$D$_1  $  &   3.087$-$09  & 2.783$-$09   &  2.840$-$09  &             & (3.3 $\pm$ 0.3)$-$09   &  (4.0 $\pm$ 0.4)$-$09   \\  
   25   &  3s4d     &   $^3$D$_2  $  &   3.096$-$09  & 2.797$-$09   &  2.855$-$09  &             & (3.3 $\pm$ 0.3)$-$09   &  (4.0 $\pm$ 0.4)$-$09   \\  
   26   &  3s4d     &   $^3$D$_3  $  &   3.110$-$09  & 2.818$-$09   &  2.876$-$09  &             & (3.3 $\pm$ 0.3)$-$09   &  (4.0 $\pm$ 0.4)$-$09   \\  
   27   &  3s4f     &   $^1$F$_3^o$  &   6.596$-$10  & 5.979$-$10   &  6.063$-$10  &             &                                                  \\
   28   &  3p3d     &   $^1$D$_2^o$  &   4.102$-$10  & 4.264$-$10   &  4.335$-$10  & 3.80$-$10   &                                                  \\
   29   &  3s4d     &   $^1$D$_2  $  &   1.078$-$09  & 1.294$-$09   &  1.323$-$09  &             & (1.25 $\pm$ 0.15)$-$09 &  (1.9 $\pm$ 0.3)$-$09   \\
\hline   
\end{longtable}
\begin{flushleft}
GRASP1: present calculations from the {\sc grasp} code for 141 levels \\ 
MCHF1:  {\em ab intio} calculations of \cite{cff} with the {\sc mchf} code and available on the website: {\tt  http://nlte.nist.gov/MCHF/view.html} \\  
                
MCHF2:  {\em adjusted energy} calculations of \cite{cff} with the {\sc mchf} code and available on the website: {\tt  http://nlte.nist.gov/MCHF/view.html} \\
MBPT: calculations of \cite{saf} with the {\sc mbpt} code \\
Expt.1: measurements of \cite{bash} \\ 
Expt.2: measurements of \cite{berry} \\ 
Expt.3: measurements of \cite{hsk} \\ 
Expt.4: measurements of \cite{liva} \\ 
Expt.5: measurements of \cite{livb} \\ 
\end{flushleft}

\clearpage


\begin{longtable}{rrrrrrrrrrrrr}
\caption{\label{table_D}
Comparison of collision strengths ($\Omega$) for  some  transitions of Si III. 
}
I &
\multicolumn{1}{r}{J} &
\multicolumn{1}{l}{Transition} &
\multicolumn{1}{c}{RM} &
\multicolumn{4}{c}{DARC} &
\multicolumn{1}{c}{FAC} \\
\hline
\multicolumn{2}{l}{Energy (Ryd)} & 
\multicolumn{1}{l}{} &
\multicolumn{1}{l}{0.7 -- 10} &
\multicolumn{1}{l}{4} &
\multicolumn{1}{l}{6} &
\multicolumn{1}{l}{8} &
\multicolumn{1}{l}{10} &
\multicolumn{1}{l}{10} \\
\hline
\endfirsthead
\caption[]{(continued)}
I &
\multicolumn{1}{r}{J} &
\multicolumn{1}{l}{Transition} &
\multicolumn{1}{c}{RM} &
\multicolumn{4}{c}{DARC} &
\multicolumn{1}{c}{FAC} \\
\hline
\multicolumn{2}{l}{Energy (Ryd)} & 
\multicolumn{1}{l}{} &
\multicolumn{1}{l}{0.7 -- 10} &
\multicolumn{1}{l}{4} &
\multicolumn{1}{l}{6} &
\multicolumn{1}{l}{8} &
\multicolumn{1}{l}{10} &
\multicolumn{1}{l}{10} \\
\hline
\endhead
1  &  6  &  3s$^2$~$^1$S$_0$ -- 3p$^2$~$^1$D$_2$ &  1.21   &  1.3780  &  1.6062  &  1.7271  &  1.8087  &  1.9381  \\
1  &  9  &  3s$^2$~$^1$S$_0$ -- 3p$^2$~$^3$P$_2$ &  0.013  &  0.0025  &  0.0015  &  0.0010  &  0.0009  &  0.0006  \\
1  & 10  &  3s$^2$~$^1$S$_0$ -- 3s3d~$^3$D$_3$   &  0.275  &  0.1102  &  0.0569  &  0.0342  &  0.0226  &  0.0141  \\
1  & 13  &  3s$^2$~$^1$S$_0$ -- 3s4s~$^3$S$_1$   &  0.062  &  0.0193  &  0.0083  &  0.0048  &  0.0032  &  0.0027  \\
1  & 14  &  3s$^2$~$^1$S$_0$ -- 3p$^2$~$^1$S$_0$ &  0.065  &  0.3069  &  0.3386  &  0.3498  &  0.3525  &  0.1805  \\
1  & 15  &  3s$^2$~$^1$S$_0$ -- 3s4s~$^1$S$_0$   &  0.611  &  0.5614  &  0.6180  &  0.6568  &  0.6863  &  1.0634  \\
1  & 16  &  3s$^2$~$^1$S$_0$ -- 3s3d~$^1$D$_2$   &  1.40   &  1.1635  &  1.3106  &  1.3810  &  1.4186  &  1.1485  \\
1  & 19  &  3s$^2$~$^1$S$_0$ -- 3s3p~$^3$P$^o_2$  &  0.053  &  0.0232  &  0.0109  &  0.0065  &  0.0043  &  0.0028  \\
1  & 20  &  3s$^2$~$^1$S$_0$ -- 3s3p~$^1$P$^o_1$  &  0.295  &  0.3159  &  0.3505  &  0.3712  &  0.3875  &  0.4266  \\
\hline									 	
\end{longtable}

\begin{flushleft}
RM: $R$-matrix calculations of \cite{dk1} for 20 levels \\
DARC: present calculations with the {\sc darc} code for 141 levels \\ 
FAC: present calculations with the {\sc fac} code for 141 levels \\
\end{flushleft}



\begin{longtable}{rrrrrrrrrr}
\caption{\label{table_E}
Comparison of effective collision strengths ($\Upsilon$) for  some  transitions of Si III at a temperature of 45~000~K. $a{\pm}b \equiv a{\times}$10$^{{\pm}b}$.
}
I &
\multicolumn{1}{r}{J} &
\multicolumn{1}{l}{Transition} &
\multicolumn{1}{l}{DARC}  &
\multicolumn{1}{l}{RM} &
\multicolumn{1}{l}{ICFT1a} &
\multicolumn{1}{l}{ICFT1b} &
\multicolumn{1}{l}{ICFT2} \\
\hline
\endfirsthead
\caption[]{(continued)}
I &
\multicolumn{1}{r}{J} &
\multicolumn{1}{l}{Transition} &
\multicolumn{1}{l}{DARC}  &
\multicolumn{1}{l}{RM} &
\multicolumn{1}{l}{ICFT1a} &
\multicolumn{1}{l}{ICFT1b} &
\multicolumn{1}{l}{ICFT2} \\
\hline
\endhead
1  &  6  &  3s$^2$~$^1$S$_0$ -- 3p$^2$~$^1$D$_2$ &  8.999$-$1  &  1.00$-$0  &  9.74$-$1  &  9.75$-$1  & 9.10$-$1 \\
1  &  9  &  3s$^2$~$^1$S$_0$ -- 3p$^2$~$^3$P$_2$ &  7.142$-$2  &  8.10$-$2  &  7.76$-$2  &  7.73$-$2  & 6.54$-$2 \\
1  & 10  &  3s$^2$~$^1$S$_0$ -- 3s3d~$^3$D$_3$   &  4.080$-$1  &  4.55$-$1  &  4.06$-$1    &  4.06$-$1  & 3.95$-$1 \\
1  & 13  &  3s$^2$~$^1$S$_0$ -- 3s4s~$^3$S$_1$   &  2.207$-$1  &  1.85$-$1  &            &  2.54$-$1  & 2.11$-$1 \\
1  & 14  &  3s$^2$~$^1$S$_0$ -- 3p$^2$~$^1$S$_0$ &  2.236$-$1  &  5.98$-$2  &  1.81$-$1  &  1.80$-$1  & 1.11$-$1 \\
1  & 15  &  3s$^2$~$^1$S$_0$ -- 3s4s~$^1$S$_0$   &  4.498$-$1  &  5.10$-$1  &  1.28$-$2  &  4.46$-$1  & 4.49$-$1 \\
1  & 16  &  3s$^2$~$^1$S$_0$ -- 3s3d~$^1$D$_2$   &  7.653$-$1  &  1.11$-$0  &  1.72$-$1  &  9.42$-$1  & 8.33$-$1 \\
1  & 19  &  3s$^2$~$^1$S$_0$ -- 3s3p~$^3$P$^o_2$ &  1.235$-$1  &  9.41$-$2  &            &  1.06$-$1  & 1.02$-$1 \\
1  & 20  &  3s$^2$~$^1$S$_0$ -- 3s3p~$^1$P$^o_1$ &  2.258$-$1  &  2.58$-$1  &            &  2.06$-$1  & 2.03$-$1 \\
\hline									 	
\end{longtable}
\begin{flushleft}                                                                                                              
DARC: present calculations from the {\sc darc} code for 141 levels \\ 
RM: $R$-matrix calculations of Dufton \& Kingston (1989) for 20 levels \\
ICFT1a: calculations of Griffin et al. (1999) with the {\sc icft} code for 45 levels \\
ICFT1b: calculations of Griffin et al. (1999) with the {\sc icft} code for 45 levels  available at the website:  {\tt http://www.open.adas.ac.uk} \\
ICFT2: calculations of \cite{icft1} with the {\sc icft} code for 283 levels   and available on the website:  {\tt http://amdpp.phys.strath.ac.uk/UK\_APAP/DATA/adf04/}  \\
                                                                                                                             
\end{flushleft} 
\clearpage

\TableExplanation

\renewcommand{\arraystretch}{1.0}

\section*{Table 1.\label{tbl_tab1} Comparison of energy levels  of Si~III (in Ryd).}
\begin{tabular}{@{}p{1in}p{6in}@{}}
Index            & Level index \\
Configuration    & The configuration to which the level belongs \\
Level            & The $LSJ$ designation of the level \\
NIST             & Energies compiled by NIST    and available on the website: {\tt http://www.nist.gov/pml/data/asd.cfm} \\
GRASP1a     & present calculations from the {\sc grasp} code for 141 levels {\em without} Breit and QED effects \\ 
GRASP1b     & present calculations from the {\sc grasp} code for 141 levels {\em with} Breit and QED effects \\
GRASP2       & present calculations from the {\sc grasp} code for 283 levels {\em with} Breit and QED effects \\
FAC1            & present calculations from the {\sc fac} code for 283 levels  \\
FAC2            & present calculations from the {\sc fac} code for 1211 levels \\ 
AS                & calculations of \cite{icft1} from the {\sc as} code for 283 levels  and available on the website:  {\tt http://amdpp.phys.strath.ac.uk/UK\_APAP/DATA/adf04/} \\ 
\end{tabular}

\medskip
\section*{Table 2.\label{tbl_tab2} 
 Transition wavelengths ($\lambda_{ij}$, in $\rm \AA$), radiative rates (A$_{ji}$, in s$^{-1}$), oscillator strengths (f$_{ij}$, dimensionless), and line  
strengths (S, in atomic units) for electric dipole (E1), and A$_{ji}$ for E2, M1 and M2 transitions in Si~III. $a{\pm}b \equiv a{\times}$10$^{{\pm}b}$. See Table 1 for level indices.          }
\begin{tabular}{@{}p{1in}p{6in}@{}}
$i$ and $j$        & The lower ($i$)  and  upper ($j$) levels of a transition as 
                      defined in Table~\ref{tbl_tab1} \\
$\lambda_{ij}$           & Transition wavelength (in \AA) \\
$A_{ji}^{E1}$                 & Radiative transition probability (in s$^{-1}$)  for the E1 transitions  \\
$f_{ij}^{E1}$                & oscillator strength (dimensionless)   for the E1 transitions  \\
$S^{E1}$                 & Transition line strength $S$ in atomic unit (a.u.), 1 a.u. = 
                      6.460$\times$10$^{-36}$ cm$^2$ esu$^2$    for the E1 transitions \\
$A_{ji}^{E2}$                 & Radiative transition probability (in s$^{-1}$) for the E2 transitions  \\                      
 $A_{ji}^{M1}$                 & Radiative transition probability (in s$^{-1}$)  for the M1 transitions  \\                     
$A_{ji}^{M2}$                 & Radiative transition probability (in s$^{-1}$)  for the M2 transitions   \\                      
\end{tabular}

\medskip
\section*{Table 3.\label{tbl_tab3} 
Collision strengths ($\Omega$) for resonance transitions of  Si III. $a{\pm}b \equiv$ $a\times$10$^{{\pm}b}$.  See Table 1 for level indices.          }
\begin{tabular}{@{}p{1in}p{6in}@{}}
$i$ and $j$         & The lower ($i$)  and  upper ($j$) levels of a transition as 
                      defined in Table~\ref{tbl_tab1} \\
$\Omega$                 & Collision strengths (dimensionless) at 8 energies of 4, 6, 8, 10, 15, 20, 25 and 30~Ryd  \\
                  
\end{tabular}

\medskip
\section*{Table 4.\label{tbl_tab4} 
Effective collision strengths ($\Upsilon$) for transitions in  Si III. $a{\pm}b \equiv a{\times}10^{{\pm}b}$.  See Table 1 for level indices.          }
\begin{tabular}{@{}p{1in}p{6in}@{}}
$i$ and $j$         & The lower ($i$)  and  upper ($j$) levels of a transition as 
                      defined in Table~\ref{tbl_tab1} \\
$\Upsilon$                 & Effective collision strengths (dimensionless) at 10 electron temperatures of 4.1, 4.3, 4.5, 4.7, 4.9, 5.1, 5.3, 5.5, 5.7, and 5.9 (log, K)  \\                     
\end{tabular}


\datatables 


\setlength{\LTleft}{0pt}
\setlength{\LTright}{0pt}

\renewcommand{\arraystretch}{1.50}

\footnotesize  

\begin{longtable}{@{\extracolsep\fill}rllllllllllll@{}}
\caption{\label{tbl_tab1}
Comparison of energy levels  of Si~III (in Ryd).
}
Index &
\multicolumn{1}{c}{Configuration} &
\multicolumn{1}{c}{Level} &
\multicolumn{1}{c}{NIST} &
\multicolumn{1}{c}{GRASP1a} &
\multicolumn{1}{r}{GRASP1b}&
\multicolumn{1}{r}{GRASP2} &
\multicolumn{1}{c}{FAC1} &
\multicolumn{1}{c}{FAC2} &
\multicolumn{1}{c}{AS} \\
\hline
\endfirsthead\\
\caption[]{(continued)}
Index &
\multicolumn{1}{c}{Configuration} &
\multicolumn{1}{c}{Level} &
\multicolumn{1}{c}{NIST} &
\multicolumn{1}{c}{GRASP1a} &
\multicolumn{1}{r}{GRASP1b}&
\multicolumn{1}{r}{GRASP2} &
\multicolumn{1}{c}{FAC1} &
\multicolumn{1}{c}{FAC2} &
\multicolumn{1}{c}{AS} \\
\hline
\endhead\\
    1   &  3s$^2$   &   $^1$S$_0  $  & 0.00000 & 0.00000  & 0.00000  & 0.00000 & 0.00000  &  0.00000  &  0.00000     \\
    2   &  3s3p     &   $^3$P$_0^o$  & 0.48046 & 0.46493  & 0.46491  & 0.46603 & 0.47091  &  0.47174  &  0.46617     \\
    3   &  3s3p     &   $^3$P$_1^o$  & 0.48163 & 0.46613  & 0.46603  & 0.46715 & 0.47196  &  0.47280  &  0.46718     \\
    4   &  3s3p     &   $^3$P$_2^o$  & 0.48430 & 0.46855  & 0.46832  & 0.46945 & 0.47411  &  0.47497  &  0.46922     \\
    5   &  3s3p     &   $^1$P$_1^o$  & 0.75530 & 0.76909  & 0.76885  & 0.76640 & 0.76943  &  0.76733  &  0.76601     \\
    6   &  3p$^2$   &   $^1$D$_2  $  & 1.11370 & 1.09124  & 1.09082  & 1.09210 & 1.09960  &  1.10069  &  1.09529     \\
    7   &  3p$^2$   &   $^3$P$_0  $  & 1.18199 & 1.17041  & 1.17017  & 1.17070 & 1.17941  &  1.17975  &  1.17144     \\
    8   &  3p$^2$   &   $^3$P$_1  $  & 1.18321 & 1.17162  & 1.17133  & 1.17186 & 1.18049  &  1.18085  &  1.17247     \\
    9   &  3p$^2$   &   $^3$P$_2  $  & 1.18556 & 1.17400  & 1.17357  & 1.17411 & 1.18259  &  1.18297  &  1.17450     \\
   10   &  3s3d     &   $^3$D$_3  $  & 1.30260 & 1.30163  & 1.30103  & 1.29921 & 1.29811  &  1.29742  &  1.29634     \\
   11   &  3s3d     &   $^3$D$_2  $  & 1.30262 & 1.30163  & 1.30106  & 1.29924 & 1.29817  &  1.29748  &  1.29630     \\
   12   &  3s3d     &   $^3$D$_1  $  & 1.30264 & 1.30164  & 1.30108  & 1.29927 & 1.29822  &  1.29753  &  1.29626     \\
   13   &  3s4s     &   $^3$S$_1  $  & 1.39767 & 1.37815  & 1.37767  & 1.37810 & 1.39249  &  1.39361  &  1.38114     \\
   14   &  3p$^2$   &   $^1$S$_0  $  & 1.39829 & 1.40097  & 1.40051  & 1.40136 & 1.41018  &  1.41013  &  1.41341     \\
   15   &  3s4s     &   $^1$S$_0  $  & 1.44955 & 1.45505  & 1.45466  & 1.44717 & 1.46498  &  1.46081  &  1.45240     \\
   16   &  3s3d     &   $^1$D$_2  $  & 1.51056 & 1.55112  & 1.55057  & 1.53971 & 1.53518  &  1.52829  &  1.53391     \\
   17   &  3s4p     &   $^3$P$_0^o$  & 1.59681 & 1.57538  & 1.57494  & 1.57591 & 1.59182  &  1.59243  &  1.58566     \\
   18   &  3s4p     &   $^3$P$_1^o$  & 1.59711 & 1.57569  & 1.57523  & 1.57620 & 1.59203  &  1.59266  &  1.58594     \\
   19   &  3s4p     &   $^3$P$_2^o$  & 1.59779 & 1.57635  & 1.57585  & 1.57683 & 1.59248  &  1.59312  &  1.58653     \\
   20   &  3s4p     &   $^1$P$_1^o$  & 1.60827 & 1.59278  & 1.59231  & 1.59153 & 1.60723  &  1.60643  &  1.59734     \\
   21   &  3p3d     &   $^3$F$_2^o$  & 1.81272 & 1.79188  & 1.79124  & 1.79020 & 1.79820  &  1.79773  &  1.79591     \\
   22   &  3p3d     &   $^3$F$_3^o$  & 1.81366 & 1.79279  & 1.79210  & 1.79107 & 1.79902  &  1.79856  &  1.79670     \\
   23   &  3p3d     &   $^3$F$_4^o$  & 1.81492 & 1.79400  & 1.79323  & 1.79221 & 1.80012  &  1.79967  &  1.79774     \\
   24   &  3s4d     &   $^3$D$_1  $  & 1.83709 & 1.81677  & 1.81620  & 1.81776 & 1.82199  &  1.82298  &  1.82555     \\
   25   &  3s4d     &   $^3$D$_2  $  & 1.83710 & 1.81678  & 1.81620  & 1.81777 & 1.82197  &  1.82297  &  1.82557     \\
   26   &  3s4d     &   $^3$D$_3  $  & 1.83711 & 1.81680  & 1.81621  & 1.81777 & 1.82194  &  1.82294  &  1.82561     \\
   27   &  3s4f     &   $^1$F$_3^o$  & 1.86653 & 1.84429  & 1.84367  & 1.84361 & 1.85391  &  1.85407  &  1.85111     \\
   28   &  3p3d     &   $^1$D$_2^o$  & 1.86836 & 1.84843  & 1.84767  & 1.84832 & 1.85357  &  1.85469  &  1.85203     \\
   29   &  3s4d     &   $^1$D$_2  $  & 1.86200 & 1.86491  & 1.86430  & 1.85842 & 1.86306  &  1.85943  &  1.86439     \\
   30   &  3s4f     &   $^3$F$_2^o$  & 1.90939 & 1.89340  & 1.89277  & 1.89121 & 1.90018  &  1.90000  &  1.90018     \\
   31   &  3s4f     &   $^3$F$_3^o$  & 1.90965 & 1.89374  & 1.89309  & 1.89150 & 1.90043  &  1.90023  &  1.90045     \\
   32   &  3s4f     &   $^3$F$_4^o$  & 1.91001 & 1.89423  & 1.89354  & 1.89191 & 1.90079  &  1.90058  &  1.90082     \\
   33   &  3p3d     &   $^3$P$_2^o$  & 1.97007 & 1.96414  & 1.96340  & 1.96128 & 1.96510  &  1.96437  &  1.96489     \\
   34   &  3p3d     &   $^3$P$_1^o$  & 1.97097 & 1.96535  & 1.96458  & 1.96246 & 1.96627  &  1.96552  &  1.96577     \\
   35   &  3p3d     &   $^3$P$_0^o$  & 1.97153 & 1.96603  & 1.96524  & 1.96313 & 1.96693  &  1.96618  &  1.96628     \\
   36   &  3p3d     &   $^3$D$_1^o$  & 1.98096 & 1.97845  & 1.97772  & 1.97540 & 1.97767  &  1.97689  &  1.97409     \\
   37   &  3p3d     &   $^3$D$_2^o$  & 1.98146 & 1.97894  & 1.97817  & 1.97585 & 1.97808  &  1.97731  &  1.97453     \\
   38   &  3p3d     &   $^3$D$_3^o$  & 1.98191 & 1.97942  & 1.97861  & 1.97629 & 1.97847  &  1.97770  &  1.97495     \\
   39   &  3p4s     &   $^3$P$_0^o$  & 2.06311 & 2.03985  & 2.03934  & 2.04158 & 2.06339  &  2.06366  &  2.05752     \\
   40   &  3p4s     &   $^3$P$_1^o$  & 2.06327 & 2.04108  & 2.04051  & 2.04276 & 2.06448  &  2.06469  &  2.05842     \\
   41   &  3p4s     &   $^3$P$_2^o$  & 2.06694 & 2.04376  & 2.04306  & 2.04526 & 2.06675  &  2.06699  &  2.06027     \\
   42   &  3p4s     &   $^1$P$_1^o$  & 2.08407 & 2.08464  & 2.08402  & 2.09773 & 2.12747  &  2.08292  &  2.12065     \\
   43   &  3p3d     &   $^1$F$_3^o$  & 2.14525 & 2.18534  & 2.18457  & 2.20377 & 2.20287  &           &  2.20702     \\
   44   &  3p3d     &   $^1$P$_1^o$  & 2.15014 & 2.18598  & 2.18522  & 2.17324 & 2.17197  &  2.16384  &  2.17717     \\
   45   &  3p4p     &   $^1$P$_1  $  & 2.21333 & 2.18680  & 2.18612  & 2.18693 & 2.20251  &  2.20399  &  2.19865     \\
   46   &  3p4p     &   $^3$D$_1  $  & 2.23021 & 2.20258  & 2.20199  & 2.20624 & 2.22317  &  2.22090  &  2.22118     \\
   47   &  3p4p     &   $^3$D$_2  $  & 2.23138 & 2.20410  & 2.20344  & 2.20769 & 2.22459  &  2.22189  &  2.22238     \\
   48   &  3p4p     &   $^3$D$_3  $  & 2.23339 & 2.20668  & 2.20591  & 2.21010 & 2.22666  &  2.22334  &  2.22428     \\
   49   &  3p4p     &   $^3$P$_0  $  & 2.23195 & 2.23441  & 2.23387  & 2.23444 & 2.26013  &  2.26000  &  2.26621     \\
   50   &  3p4p     &   $^3$P$_1  $  & 2.23200 & 2.23518  & 2.23460  & 2.23518 & 2.26015  &  2.26220  &  2.26690     \\
   51   &  3p4p       &   $^3$P$_2  $  & 2.23209 & 2.23696  & 2.23629  & 2.23688 & 2.26233  &  2.26419  &  2.26799   \\    
   52   &  3p4p       &   $^3$S$_1  $  & 2.26699 & 2.24617  & 2.24546  & 2.24630 & 2.26246  &  2.26591  &  2.25977   \\
   53   &  3p4p       &   $^1$D$_2  $  & 2.25935 & 2.28199  & 2.28137  & 2.28769 & 2.31315  &           &  2.31166   \\
   54   &  3p4p       &   $^1$S$_0  $  & 2.36000 & 2.36378  & 2.36320  & 2.34289 & 2.36962  &  2.36323  &  2.37654   \\    
   55   &  3p4d       &   $^1$D$_2^o$  & 2.43749 & 2.44106  & 2.44031  & 2.44219 & 2.45291  &  2.45470  &  2.46804   \\
   56   &  3p4d       &   $^3$F$_2^o$  &         & 2.44839  & 2.44769  & 2.44941 & 2.46175  &  2.46365  &  2.48034   \\
   57   &  3p4d       &   $^3$F$_3^o$  &         & 2.44947  & 2.44873  & 2.45045 & 2.46129  &  2.46319  &  2.48123   \\
   58   &  3p4d       &   $^3$F$_4^o$  & 2.43934 & 2.45155  & 2.45070  & 2.45242 & 2.46429  &  2.46620  &  2.48226   \\   
   59   &  3p4d       &   $^3$D$_1^o$  &         & 2.45784  & 2.45711  & 2.45864 & 2.46884  &  2.47000  &  2.47729   \\
   60   &  3p4d       &   $^3$D$_2^o$  &         & 2.45841  & 2.45764  & 2.45917 & 2.46823  &  2.46940  &  2.47741   \\
   61   &  3p4d       &   $^3$D$_3^o$  &         & 2.45920  & 2.45838  & 2.45991 & 2.46962  &  2.47081  &  2.47787   \\
   62   &  3p4d       &   $^3$P$_2^o$  &         & 2.48056  & 2.47981  & 2.48031 & 2.49131  &  2.49178  &  2.50453   \\
   63   &  3p4d       &   $^3$P$_1^o$  &         & 2.48183  & 2.48103  & 2.48151 & 2.49260  &  2.49306  &  2.50537   \\
   64   &  3p4d       &   $^3$P$_0^o$  &         & 2.48247  & 2.48165  & 2.48211 & 2.49262  &  2.49309  &  2.50580   \\
   65   &  3p4f       &   $^3$F$_4  $  &         & 2.48375  & 2.48298  & 2.48419 & 2.49923  &  2.50027  &  2.50781   \\
   66   &  3p4f       &   $^3$F$_2  $  &         & 2.48414  & 2.48336  & 2.48407 & 2.49886  &  2.49964  &  2.50777   \\
   67   &  3p4f       &   $^3$F$_3  $  &         & 2.48443  & 2.48363  & 2.48435 & 2.49944  &  2.50023  &  2.50798   \\
   68   &  3p4f       &   $^1$G$_4  $  &         & 2.48610  & 2.48521  & 2.48672 & 2.50105  &  2.50381  &  2.50941   \\
   69   &  3p4f       &   $^1$F$_3  $  &         & 2.49180  & 2.49103  & 2.49203 & 2.50706  &  2.50809  &  2.51659   \\
   70   &  3p4f       &   $^3$G$_3  $  &         & 2.50475  & 2.50402  & 2.50713 & 2.52138  &  2.52568  &  2.53226   \\
   71   &  3p4f       &   $^3$G$_4  $  &         & 2.50610  & 2.50531  & 2.50842 & 2.52314  &  2.52745  &  2.53318   \\
   72   &  3p4f       &   $^3$G$_5  $  &         & 2.50768  & 2.50682  & 2.50994 & 2.52404  &  2.52835  &  2.53429   \\
   73   &  3p4d       &   $^1$F$_3^o$  &         & 2.52039  & 2.51958  & 2.51535 & 2.52379  &  2.52724  &  2.52616   \\
   74   &  3p4f       &   $^1$D$_2  $  &         & 2.52589  & 2.52509  & 2.52479 & 2.53838  &  2.53886  &  2.54964   \\
   75   &  3p4f       &   $^3$D$_3  $  &         & 2.52883  & 2.52808  & 2.52720 & 2.54190  &  2.54225  &  2.55412   \\
   76   &  3p4f       &   $^3$D$_2  $  &         & 2.53042  & 2.52959  & 2.52870 & 2.54303  &  2.54338  &  2.55508   \\
   77   &  3p4f       &   $^3$D$_1  $  &         & 2.53105  & 2.53021  & 2.52925 & 2.54397  &  2.54431  &  2.55557   \\
   78   &  3p4d       &   $^1$P$_1^o$  &         & 2.56389  & 2.56313  & 2.54271 & 2.55313  &  2.54705  &  2.55396   \\
   79   &  3d$^2$     &   $^3$F$_2  $  &         & 2.68715  & 2.68601  & 2.66494 & 2.67322  &  2.67008  &  2.67839   \\
   80   &  3d$^2$     &   $^3$F$_3  $  &         & 2.68717  & 2.68601  & 2.66506 & 2.67335  &  2.67021  &  2.67850   \\
   81   &  3d$^2$     &   $^3$F$_4  $  &         & 2.68720  & 2.68601  & 2.66518 & 2.67337  &  2.67023  &  2.67864   \\
   82   &  3d$^2$     &   $^3$P$_0  $  &         & 2.75379  & 2.75263  & 2.75259 & 2.75556  &  2.75716  &  2.75707   \\
   83   &  3d$^2$     &   $^3$P$_1  $  &         & 2.75379  & 2.75263  & 2.75260 & 2.75555  &  2.75716  &  2.75710   \\
   84   &  3d$^2$     &   $^3$P$_2  $  &         & 2.75381  & 2.75262  & 2.75258 & 2.75552  &  2.75715  &  2.75714   \\
   85   &  3d$^2$     &   $^1$D$_2  $  &         & 2.77710  & 2.77599  & 2.80829 & 2.81150  &           &  2.81063   \\
   86   &  3d$^2$     &   $^1$G$_4  $  &         & 2.80606  & 2.80491  & 2.85366 & 2.85931  &           &  2.87269   \\
   87   &  3d4s       &   $^3$D$_1  $  &         & 2.83872  & 2.83767  & 2.84176 & 2.85939  &  2.85216  &  2.87157   \\
   88   &  3d4s       &   $^3$D$_2  $  &         & 2.83873  & 2.83766  & 2.84177 & 2.85934  &           &  2.87161   \\
   89   &  3d4s       &   $^3$D$_3  $  &         & 2.83875  & 2.83765  & 2.84179 & 2.85928  &  2.85359  &  2.87168   \\
   90   &  3d4s       &   $^1$D$_2  $  &         & 2.88996  & 2.88887  & 2.90281 & 2.93295  &           &  2.96197   \\
   91   &  3d$^2$     &   $^1$S$_0  $  &         & 2.94903  & 2.94789  & 2.94408 & 2.94987  &  2.95116  &  2.95683   \\
   92   &  3d4p       &   $^1$D$_2^o$  &         & 2.99386  & 2.99275  & 2.99431 & 3.00824  &  3.01031  &  3.02972   \\
   93   &  3d4p       &   $^3$D$_1^o$  &         & 3.00925  & 3.00817  & 3.01273 & 3.03227  &  3.04207  &  3.05616   \\
   94   &  3d4p       &   $^3$D$_2^o$  &         & 3.00936  & 3.00826  & 3.01282 & 3.03031  &  3.03401  &  3.05629   \\
   95   &  3d4p       &   $^3$D$_3^o$  &         & 3.00954  & 3.00842  & 3.01298 & 3.02958  &  3.03335  &  3.05647   \\
   96   &  3d4p       &   $^3$F$_2^o$  &         & 3.01719  & 3.01612  & 3.01899 & 3.03172  &  3.04156  &  3.04809   \\
   97   &  3d4p       &   $^3$F$_3^o$  &         & 3.01757  & 3.01647  & 3.01934 & 3.03224  &  3.04203  &  3.04843   \\
   98   &  3d4p       &   $^3$F$_4^o$  &         & 3.01807  & 3.01693  & 3.01980 & 3.03067  &  3.03439  &  3.04891   \\
   99   &  3d4p       &   $^3$P$_2^o$  &         & 3.05145  & 3.05034  & 3.05280 & 3.06787  &  3.07060  &  3.08826   \\
  100   &  3d4p       &   $^3$P$_1^o$  &         & 3.05175  & 3.05066  & 3.05312 & 3.06833  &  3.07106  &  3.08850   \\   
  101   &  3d4p       &   $^3$P$_0^o$  &  & 3.05191  & 3.05082  & 3.05328 & 3.06812  & 3.07085  &  3.08862   \\    
  102   &  3d4p       &   $^1$F$_3^o$  &  & 3.07416  & 3.07308  & 3.08415 & 3.10993  & 3.13158  &  3.13350   \\    
  103   &  3d4p       &   $^1$P$_1^o$  &  & 3.10444  & 3.10340  & 3.11166 & 3.13401  & 3.15623  &  3.14704   \\    
  104   &  3d4d       &   $^1$F$_3  $  &  & 3.22055  & 3.21933  & 3.21952 & 3.21767  & 3.21835  &  3.24330   \\    
  105   &  3d4d       &   $^3$D$_1  $  &  & 3.23131  & 3.23010  & 3.23101 & 3.22990  & 3.23111  &  3.25538   \\    
  106   &  3d4d       &   $^3$D$_2  $  &  & 3.23131  & 3.23009  & 3.23101 & 3.22931  & 3.23052  &  3.25541   \\    
  107   &  3d4d       &   $^3$D$_3  $  &  & 3.23131  & 3.23008  & 3.23100 & 3.22963  & 3.23083  &  3.25545   \\    
  108   &  3d4d       &   $^1$P$_1  $  &  & 3.23834  & 3.23712  & 3.23840 & 3.23768  & 3.23902  &  3.26374   \\    
  109   &  3d4d       &   $^3$G$_3  $  &  & 3.23881  & 3.23761  & 3.23882 & 3.23827  & 3.23955  &  3.26406   \\    
  110   &  3d4d       &   $^3$G$_4  $  &  & 3.23882  & 3.23760  & 3.23881 & 3.23747  & 3.23875  &  3.26411   \\    
  111   &  3d4d       &   $^3$G$_5  $  &  & 3.23883  & 3.23758  & 3.23879 & 3.23796  & 3.23924  &  3.26418   \\    
  112   &  3d4d       &   $^3$S$_1  $  &  & 3.26075  & 3.25953  & 3.26169 & 3.26205  & 3.26402  &  3.29012   \\    
  113   &  3d4d       &   $^3$F$_4  $  &  & 3.26738  & 3.26615  & 3.26799 & 3.28500  & 3.28700  &  3.33364   \\    
  114   &  3d4d       &   $^3$F$_3  $  &  & 3.26738  & 3.26617  & 3.26801 & 3.28461  & 3.28661  &  3.33367   \\    
  115   &  3d4d       &   $^3$F$_2  $  &  & 3.26739  & 3.26619  & 3.26803 & 3.28528  & 3.28728  &  3.33371   \\    
  116   &  3d4f       &   $^1$G$_4^o$  &  & 3.27350  & 3.27227  & 3.27276 & 3.27845  & 3.27871  &  3.30982   \\    
  117   &  3d4f       &   $^3$H$_4^o$  &  & 3.28202  & 3.28080  & 3.28256 & 3.28532  & 3.28702  &  3.31445   \\    
  118   &  3d4f       &   $^3$H$_5^o$  &  & 3.28202  & 3.28078  & 3.28256 & 3.28582  & 3.28752  &  3.31449   \\    
  119   &  3d4f       &   $^3$H$_6^o$  &  & 3.28202  & 3.28077  & 3.28255 & 3.28530  & 3.28702  &  3.31453   \\    
  120   &  3d4f       &   $^3$F$_2^o$  &  & 3.28752  & 3.28630  & 3.28794 & 3.29414  & 3.29543  &  3.32719   \\    
  121   &  3d4f       &   $^3$F$_3^o$  &  & 3.28753  & 3.28630  & 3.28794 & 3.29450  & 3.29578  &  3.32721   \\    
  122   &  3d4f       &   $^3$F$_4^o$  &  & 3.28754  & 3.28630  & 3.28794 & 3.29421  & 3.29549  &  3.32723   \\    
  123   &  3d4f       &   $^1$D$_2^o$  &  & 3.30258  & 3.30134  & 3.30419 & 3.31125  & 3.31274  &  3.34706   \\    
  124   &  3d4d       &   $^3$P$_0  $  &  & 3.30652  & 3.30533  & 3.30328 & 3.32200  & 3.32187  &  3.36788   \\    
  125   &  3d4d       &   $^3$P$_1  $  &  & 3.30652  & 3.30533  & 3.30359 & 3.32126  & 3.32113  &  3.36789   \\    
  126   &  3d4d       &   $^3$P$_2  $  &  & 3.30653  & 3.30532  & 3.30359 & 3.32150  & 3.32138  &  3.36792   \\    
  127   &  3d4d       &   $^1$D$_2  $  &  & 3.30703  & 3.30584  & 3.30358 & 3.32238  & 3.32366  &  3.36529   \\    
  128   &  3d4d       &   $^1$G$_4  $  &  & 3.30785  & 3.30664  & 3.30541 & 3.32341  & 3.32600  &  3.37048   \\    
  129   &  3d4f       &   $^3$G$_3^o$  &  & 3.31549  & 3.31426  & 3.31694 & 3.33103  & 3.33394  &  3.37398   \\    
  130   &  3d4f       &   $^3$G$_4^o$  &  & 3.31549  & 3.31425  & 3.31693 & 3.33146  & 3.33438  &  3.37400   \\    
  131   &  3d4f       &   $^3$G$_5^o$  &  & 3.31549  & 3.31424  & 3.31692 & 3.33105  & 3.33397  &  3.37403   \\    
  132   &  3d4f       &   $^3$D$_1^o$  &  & 3.32947  & 3.32824  & 3.32884 & 3.34263  & 3.34444  &  3.38580   \\    
  133   &  3d4f       &   $^3$D$_2^o$  &  & 3.32948  & 3.32824  & 3.32885 & 3.34288  & 3.34469  &  3.38580   \\    
  134   &  3d4f       &   $^3$D$_3^o$  &  & 3.32948  & 3.32824  & 3.32885 & 3.34273  & 3.34454  &  3.38581   \\    
  135   &  3d4f       &   $^3$P$_2^o$  &  & 3.33868  & 3.33746  & 3.33883 & 3.34932  & 3.35154  &  3.38935   \\    
  136   &  3d4f       &   $^3$P$_1^o$  &  & 3.33869  & 3.33746  & 3.33883 & 3.34920  & 3.35143  &  3.38938   \\    
  137   &  3d4f       &   $^3$P$_0^o$  &  & 3.33870  & 3.33746  & 3.33884 & 3.34943  & 3.35166  &  3.38940   \\    
  138   &  3d4f       &   $^1$F$_3^o$  &  & 3.34011  & 3.33888  & 3.33677 & 3.35318  & 3.35537  &  3.40009   \\    
  139   &  3d4f       &   $^1$H$_5^o$  &  & 3.36941  & 3.36816  & 3.36529 & 3.38543  & 3.38772  &  3.43628   \\    
  140   &  3d4f       &   $^1$P$_1^o$  &  & 3.38789  & 3.38667  & 3.36995 & 3.38510  & 3.38548  &  3.42938   \\    
  141   &  3d4d       &   $^1$S$_0  $  &  & 3.42671  & 3.42560  & 3.37590 & 3.39343  & 3.38682  &  3.43476   \\   
\hline\\                                                                               
\end{longtable}

\clearpage

\setlength{\LTleft}{0pt}
\setlength{\LTright}{0pt}

\renewcommand{\arraystretch}{1.1}

\footnotesize 

\begin{longtable}{@{\extracolsep\fill}rrrrlrrlll@{}}
\caption{\label{tbl_tab2}
 Transition wavelengths ($\lambda_{ij}$, in $\rm \AA$), radiative rates (A$_{ji}$, in s$^{-1}$), oscillator strengths (f$_{ij}$, dimensionless), and line  
strengths (S, in atomic units) for electric dipole (E1), and A$_{ji}$ for E2, M1 and M2 transitions in Si~III. $a{\pm}b \equiv a{\times}$10$^{{\pm}b}$.
See page \pageref{tbl_tab1} for Explanation of Tables and Table~\ref{tbl_tab1} 
for definition of level indices.
}
$i$ &
$j$ &
\multicolumn{1}{c}{$\lambda_{ij}$ (\AA)} &
\multicolumn{1}{c}{$A_{ji}^{E1}$} &
\multicolumn{1}{c}{$f_{ij}^{E1}$} &
\multicolumn{1}{c}{$S^{E1}$} &
\multicolumn{1}{c}{$A_{ji}^{E2}$} &
\multicolumn{1}{c}{$A_{ji}^{M1}$} &
\multicolumn{1}{c}{$A_{ji}^{M2}$} \\
\hline
\endfirsthead\\
\caption[]{(continued)}
$i$ &
$j$ &
\multicolumn{1}{c}{$\lambda_{ij}$ (\AA)} &
\multicolumn{1}{c}{$A_{ji}^{E1}$} &
\multicolumn{1}{c}{$f_{ij}^{E1}$} &
\multicolumn{1}{c}{$S^{E1}$} &
\multicolumn{1}{c}{$A_{ji}^{E2}$} &
\multicolumn{1}{c}{$A_{ji}^{M1}$} &
\multicolumn{1}{c}{$A_{ji}^{M2}$} \\
\hline\\
\endhead
    1 &    3 &  1.955$+$03 &  1.058$+$04 &  1.820$-$05 &  1.172$-$04 &  0.000$+$00 &  0.000$+$00 &  0.000$+$00 \\
    1 &    4 &  1.946$+$03 &  0.000$+$00 &  0.000$+$00 &  0.000$+$00 &  0.000$+$00 &  0.000$+$00 &  1.089$-$02 \\
    1 &    5 &  1.185$+$03 &  2.658$+$09 &  1.679$+$00 &  6.552$+$00 &  0.000$+$00 &  0.000$+$00 &  0.000$+$00 \\
    1 &    6 &  8.354$+$02 &  0.000$+$00 &  0.000$+$00 &  0.000$+$00 &  1.604$+$04 &  0.000$+$00 &  0.000$+$00 \\
    1 &    8 &  7.780$+$02 &  0.000$+$00 &  0.000$+$00 &  0.000$+$00 &  0.000$+$00 &  2.086$-$02 &  0.000$+$00 \\
    1 &    9 &  7.765$+$02 &  0.000$+$00 &  0.000$+$00 &  0.000$+$00 &  7.321$+$00 &  0.000$+$00 &  0.000$+$00 \\
    1 &   11 &  7.004$+$02 &  0.000$+$00 &  0.000$+$00 &  0.000$+$00 &  4.147$-$03 &  0.000$+$00 &  0.000$+$00 \\
    1 &   12 &  7.004$+$02 &  0.000$+$00 &  0.000$+$00 &  0.000$+$00 &  0.000$+$00 &  1.881$-$05 &  0.000$+$00 \\
    1 &   13 &  6.615$+$02 &  0.000$+$00 &  0.000$+$00 &  0.000$+$00 &  0.000$+$00 &  2.300$-$04 &  0.000$+$00 \\
    1 &   16 &  5.877$+$02 &  0.000$+$00 &  0.000$+$00 &  0.000$+$00 &  9.491$+$04 &  0.000$+$00 &  0.000$+$00 \\
    1 &   18 &  5.785$+$02 &  2.788$+$05 &  4.196$-$05 &  7.992$-$05 &  0.000$+$00 &  0.000$+$00 &  0.000$+$00 \\
    1 &   19 &  5.783$+$02 &  0.000$+$00 &  0.000$+$00 &  0.000$+$00 &  0.000$+$00 &  0.000$+$00 &  7.574$-$02 \\
    1 &   20 &  5.723$+$02 &  2.429$+$08 &  3.579$-$02 &  6.742$-$02 &  0.000$+$00 &  0.000$+$00 &  0.000$+$00 \\
    1 &   21 &  5.087$+$02 &  0.000$+$00 &  0.000$+$00 &  0.000$+$00 &  0.000$+$00 &  0.000$+$00 &  2.841$-$05 \\
    1 &   24 &  5.017$+$02 &  0.000$+$00 &  0.000$+$00 &  0.000$+$00 &  0.000$+$00 &  1.478$-$05 &  0.000$+$00 \\
    1 &   25 &  5.017$+$02 &  0.000$+$00 &  0.000$+$00 &  0.000$+$00 &  3.837$-$03 &  0.000$+$00 &  0.000$+$00 \\
\hline
\end{longtable}

\clearpage 

\setlength{\LTleft}{0pt}
\setlength{\LTright}{0pt}

\renewcommand{\arraystretch}{1.1}

\footnotesize 

\begin{longtable}{@{\extracolsep\fill}rrrrlrrlll@{}}
\caption{\label{tbl_tab3}
Collision strengths ($\Omega$) for resonance transitions of  Si III. $a{\pm}b \equiv$ $a\times$10$^{{\pm}b}$.
See page \pageref{tbl_tab1} for Explanation of Tables and Table~\ref{tbl_tab1} 
for definition of level indices.
}
Transition & & Energy \\
\hline
$i$  &
$j$ &
4 & 6 & 8 & 10 & 15 & 20 & 25 & 30 \\
\hline
\endfirsthead\\
\caption[]{(continued)}
Transition & & Energy \\
\hline
$i$ &
$j$ &
4 & 6 & 8 & 10 & 15 & 20 & 25 & 30 \\
\hline\\
\endhead
  1 &  2 &  2.389$-$02 &  1.367$-$02 &  8.809$-$03 &  6.093$-$03 &  2.959$-$03 &  1.729$-$03 &  1.131$-$03 &  7.978$-$04 \\
  1 &  3 &  7.212$-$02 &  4.153$-$02 &  2.698$-$02 &  1.895$-$02 &  9.660$-$03 &  6.056$-$03 &  4.342$-$03 &  3.417$-$03 \\
  1 &  4 &  1.193$-$01 &  6.820$-$02 &  4.394$-$02 &  3.038$-$02 &  1.475$-$02 &  8.616$-$03 &  5.636$-$03 &  3.974$-$03 \\
  1 &  5 &  1.636$+$01 &  2.034$+$01 &  2.292$+$01 &  2.583$+$01 &  3.017$+$01 &  3.351$+$01 &  3.637$+$01 &  3.898$+$01 \\
  1 &  6 &  1.378$+$00 &  1.606$+$00 &  1.727$+$00 &  1.809$+$00 &  1.917$+$00 &  1.983$+$00 &  2.038$+$00 &  2.093$+$00 \\
  1 &  7 &  4.376$-$04 &  2.065$-$04 &  1.162$-$04 &  7.479$-$05 &  3.699$-$05 &  2.521$-$05 &  2.003$-$05 &  1.720$-$05 \\
  1 &  8 &  1.251$-$03 &  5.518$-$04 &  2.808$-$04 &  1.588$-$04 &  5.254$-$05 &  2.333$-$05 &  1.241$-$05 &  7.371$-$06 \\
  1 &  9 &  2.524$-$03 &  1.464$-$03 &  1.047$-$03 &  8.681$-$04 &  7.136$-$04 &  6.813$-$04 &  6.780$-$04 &  6.851$-$04 \\
  1 & 10 &  1.102$-$01 &  5.687$-$02 &  3.423$-$02 &  2.260$-$02 &  1.020$-$02 &  5.702$-$03 &  3.617$-$03 &  2.494$-$03 \\
  1 & 11 &  7.874$-$02 &  4.063$-$02 &  2.445$-$02 &  1.614$-$02 &  7.289$-$03 &  4.073$-$03 &  2.584$-$03 &  1.781$-$03 \\
  1 & 12 &  4.725$-$02 &  2.437$-$02 &  1.467$-$02 &  9.685$-$03 &  4.373$-$03 &  2.443$-$03 &  1.550$-$03 &  1.069$-$03 \\
  1 & 13 &  1.939$-$02 &  8.345$-$03 &  4.835$-$03 &  3.240$-$03 &  1.578$-$03 &  9.360$-$04 &  6.186$-$04 &  4.388$-$04 \\
  1 & 14 &  3.069$-$01 &  3.386$-$01 &  3.498$-$01 &  3.525$-$01 &  3.476$-$01 &  3.398$-$01 &  3.331$-$01 &  3.277$-$01 \\
  1 & 15 &  5.614$-$01 &  6.180$-$01 &  6.568$-$01 &  6.863$-$01 &  7.351$-$01 &  7.641$-$01 &  7.831$-$01 &  7.963$-$01 \\
  1 & 16 &  1.164$+$00 &  1.311$+$00 &  1.381$+$00 &  1.419$+$00 &  1.453$+$00 &  1.464$+$00 &  1.473$+$00 &  1.487$+$00 \\
  1 & 17 &  4.646$-$03 &  2.180$-$03 &  1.301$-$03 &  8.616$-$04 &  3.915$-$04 &  2.204$-$04 &  1.411$-$04 &  9.812$-$05 \\
  1 & 18 &  1.410$-$02 &  6.731$-$03 &  4.114$-$03 &  2.812$-$03 &  1.435$-$03 &  9.492$-$04 &  7.339$-$04 &  6.248$-$04 \\
  1 & 19 &  2.324$-$02 &  1.090$-$02 &  6.510$-$03 &  4.311$-$03 &  1.958$-$03 &  1.103$-$03 &  7.056$-$04 &  4.907$-$04 \\
  1 & 20 &  3.157$-$01 &  3.505$-$01 &  3.712$-$01 &  3.875$-$01 &  4.187$-$01 &  4.423$-$01 &  4.616$-$01 &  4.780$-$01 \\
  1 & 21 &  8.398$-$03 &  3.371$-$03 &  1.709$-$03 &  9.982$-$04 &  3.834$-$04 &  2.026$-$04 &  1.268$-$04 &  8.784$-$05 \\
  1 & 22 &  1.177$-$02 &  4.725$-$03 &  2.394$-$03 &  1.397$-$03 &  5.360$-$04 &  2.839$-$04 &  1.791$-$04 &  1.257$-$04 \\
  1 & 23 &  1.513$-$02 &  6.077$-$03 &  3.073$-$03 &  1.787$-$03 &  6.762$-$04 &  3.508$-$04 &  2.154$-$04 &  1.462$-$04 \\
  1 & 24 &  1.778$-$02 &  8.301$-$03 &  4.782$-$03 &  3.088$-$03 &  1.353$-$03 &  7.437$-$04 &  4.671$-$04 &  3.195$-$04 \\
  1 & 25 &  2.961$-$02 &  1.383$-$02 &  7.968$-$03 &  5.146$-$03 &  2.256$-$03 &  1.240$-$03 &  7.788$-$04 &  5.328$-$04 \\
  1 & 26 &  4.144$-$02 &  1.935$-$02 &  1.115$-$02 &  7.201$-$03 &  3.157$-$03 &  1.735$-$03 &  1.089$-$03 &  7.452$-$04 \\
  1 & 27 &  2.418$-$01 &  2.539$-$01 &  2.586$-$01 &  2.604$-$01 &  2.612$-$01 &  2.617$-$01 &  2.635$-$01 &  2.666$-$01 \\
  1 & 28 &  3.672$-$02 &  4.064$-$02 &  4.063$-$02 &  3.922$-$02 &  3.425$-$02 &  2.972$-$02 &  2.606$-$02 &  2.314$-$02 \\
  1 & 29 &  3.914$-$01 &  4.576$-$01 &  4.904$-$01 &  5.087$-$01 &  5.294$-$01 &  5.372$-$01 &  5.409$-$01 &  5.429$-$01 \\
  1 & 30 &  6.247$-$03 &  2.212$-$03 &  1.080$-$03 &  6.284$-$04 &  2.389$-$04 &  1.228$-$04 &  7.447$-$05 &  5.009$-$05 \\
  1 & 31 &  8.730$-$03 &  3.079$-$03 &  1.498$-$03 &  8.686$-$04 &  3.273$-$04 &  1.668$-$04 &  1.007$-$04 &  6.758$-$05 \\
  1 & 32 &  1.120$-$02 &  3.930$-$03 &  1.907$-$03 &  1.102$-$03 &  4.109$-$04 &  2.065$-$04 &  1.223$-$04 &  8.029$-$05 \\
  1 & 33 &  2.226$-$03 &  1.083$-$03 &  5.732$-$04 &  3.362$-$04 &  1.208$-$04 &  5.830$-$05 &  3.368$-$05 &  2.184$-$05 \\
  1 & 34 &  1.333$-$03 &  6.480$-$04 &  3.423$-$04 &  2.004$-$04 &  7.171$-$05 &  3.471$-$05 &  2.030$-$05 &  1.348$-$05 \\
  1 & 35 &  4.451$-$04 &  2.160$-$04 &  1.138$-$04 &  6.632$-$05 &  2.331$-$05 &  1.094$-$05 &  6.116$-$06 &  3.826$-$06 \\
  1 & 36 &  4.357$-$04 &  1.763$-$04 &  8.456$-$05 &  4.643$-$05 &  1.494$-$05 &  6.756$-$06 &  3.839$-$06 &  2.574$-$06 \\
  1 & 37 &  7.305$-$04 &  2.949$-$04 &  1.407$-$04 &  7.664$-$05 &  2.377$-$05 &  1.005$-$05 &  5.144$-$06 &  2.992$-$06 \\
  1 & 38 &  1.008$-$03 &  4.060$-$04 &  1.940$-$04 &  1.060$-$04 &  3.352$-$05 &  1.475$-$05 &  8.055$-$06 &  5.137$-$06 \\
  1 & 39 &  3.631$-$04 &  1.502$-$04 &  7.177$-$05 &  3.966$-$05 &  1.333$-$05 &  6.167$-$06 &  3.418$-$06 &  2.123$-$06 \\
  1 & 40 &  1.303$-$03 &  6.372$-$04 &  3.804$-$04 &  2.675$-$04 &  1.609$-$04 &  1.229$-$04 &  1.039$-$04 &  9.249$-$05 \\
  1 & 41 &  1.793$-$03 &  7.411$-$04 &  3.537$-$04 &  1.953$-$04 &  6.555$-$05 &  3.031$-$05 &  1.679$-$05 &  1.043$-$05 \\
  1 & 42 &  1.251$-$01 &  1.064$-$01 &  9.277$-$02 &  8.263$-$02 &  6.591$-$02 &  5.600$-$02 &  4.954$-$02 &  4.505$-$02 \\
  1 & 43 &  9.107$-$02 &  8.179$-$02 &  6.916$-$02 &  5.854$-$02 &  4.084$-$02 &  3.092$-$02 &  2.487$-$02 &  2.091$-$02 \\
  1 & 44 &  1.364$-$02 &  1.858$-$02 &  1.928$-$02 &  1.870$-$02 &  1.643$-$02 &  1.471$-$02 &  1.357$-$02 &  1.282$-$02 \\
  1 & 45 &  6.487$-$03 &  5.424$-$03 &  4.540$-$03 &  3.833$-$03 &  2.649$-$03 &  1.964$-$03 &  1.530$-$03 &  1.236$-$03 \\
  1 & 46 &  6.668$-$04 &  2.156$-$04 &  1.180$-$04 &  7.696$-$05 &  3.802$-$05 &  2.436$-$05 &  1.757$-$05 &  1.355$-$05 \\
  1 & 47 &  1.040$-$03 &  2.973$-$04 &  1.462$-$04 &  8.775$-$05 &  3.985$-$05 &  2.695$-$05 &  2.189$-$05 &  1.941$-$05 \\
  1 & 48 &  1.428$-$03 &  3.889$-$04 &  1.787$-$04 &  9.757$-$05 &  3.181$-$05 &  1.459$-$05 &  8.086$-$06 &  5.031$-$06 \\
  1 & 49 &  7.126$-$05 &  2.969$-$05 &  1.461$-$05 &  8.528$-$06 &  3.714$-$06 &  2.406$-$06 &  1.870$-$06 &  1.593$-$06 \\
  1 & 50 &  2.134$-$04 &  8.630$-$05 &  4.042$-$05 &  2.190$-$05 &  7.414$-$06 &  3.631$-$06 &  2.170$-$06 &  1.460$-$06 \\                                                                                                       
  1 & 51 &  3.751$-$04 &  1.648$-$04 &  8.863$-$05 &  5.836$-$05 &  3.561$-$05 &  3.010$-$05 &  2.808$-$05 &  2.715$-$05 \\
  1 & 52 &  3.458$-$04 &  1.262$-$04 &  6.500$-$05 &  3.776$-$05 &  1.358$-$05 &  6.678$-$06 &  3.928$-$06 &  2.580$-$06 \\
  1 & 53 &  4.209$-$02 &  3.921$-$02 &  3.775$-$02 &  3.694$-$02 &  3.589$-$02 &  3.529$-$02 &  3.487$-$02 &  3.460$-$02 \\
  1 & 54 &  4.526$-$03 &  6.324$-$03 &  6.885$-$03 &  6.999$-$03 &  6.741$-$03 &  6.381$-$03 &  6.076$-$03 &  5.839$-$03 \\
  1 & 55 &  5.730$-$03 &  4.903$-$03 &  4.132$-$03 &  3.524$-$03 &  2.539$-$03 &  1.964$-$03 &  1.592$-$03 &  1.333$-$03 \\
  1 & 56 &  2.168$-$03 &  6.989$-$04 &  3.620$-$04 &  2.327$-$04 &  1.182$-$04 &  7.944$-$05 &  6.012$-$05 &  4.841$-$05 \\
  1 & 57 &  2.889$-$03 &  8.091$-$04 &  3.555$-$04 &  1.938$-$04 &  6.874$-$05 &  3.623$-$05 &  2.358$-$05 &  1.737$-$05 \\
  1 & 58 &  3.740$-$03 &  1.023$-$03 &  4.363$-$04 &  2.294$-$04 &  7.290$-$05 &  3.397$-$05 &  1.953$-$05 &  1.274$-$05 \\
  1 & 59 &  2.454$-$04 &  1.016$-$04 &  5.032$-$05 &  2.917$-$05 &  1.148$-$05 &  6.580$-$06 &  4.641$-$06 &  3.685$-$06 \\
  1 & 60 &  4.356$-$04 &  1.759$-$04 &  8.569$-$05 &  4.858$-$05 &  1.770$-$05 &  9.248$-$06 &  5.927$-$06 &  4.280$-$06 \\
  1 & 61 &  6.141$-$04 &  2.418$-$04 &  1.150$-$04 &  6.349$-$05 &  2.146$-$05 &  1.044$-$05 &  6.371$-$06 &  4.501$-$06 \\
  1 & 62 &  8.990$-$04 &  4.195$-$04 &  2.250$-$04 &  1.324$-$04 &  4.682$-$05 &  2.219$-$05 &  1.264$-$05 &  8.126$-$06 \\
  1 & 63 &  5.388$-$04 &  2.514$-$04 &  1.349$-$04 &  7.949$-$05 &  2.833$-$05 &  1.366$-$05 &  8.000$-$06 &  5.350$-$06 \\
  1 & 64 &  1.794$-$04 &  8.322$-$05 &  4.427$-$05 &  2.580$-$05 &  8.835$-$06 &  4.021$-$06 &  2.188$-$06 &  1.341$-$06 \\
  1 & 65 &  6.587$-$03 &  4.251$-$03 &  3.267$-$03 &  2.696$-$03 &  1.954$-$03 &  1.595$-$03 &  1.383$-$03 &  1.243$-$03 \\
  1 & 66 &  1.443$-$04 &  5.680$-$05 &  2.876$-$05 &  1.798$-$05 &  9.075$-$06 &  6.396$-$06 &  5.209$-$06 &  4.558$-$06 \\
  1 & 67 &  2.078$-$04 &  8.204$-$05 &  4.135$-$05 &  2.534$-$05 &  1.158$-$05 &  7.164$-$06 &  5.121$-$06 &  3.974$-$06 \\
  1 & 68 &  6.677$-$03 &  4.305$-$03 &  3.309$-$03 &  2.731$-$03 &  1.978$-$03 &  1.614$-$03 &  1.398$-$03 &  1.256$-$03 \\
  1 & 69 &  4.988$-$03 &  4.418$-$03 &  3.840$-$03 &  3.356$-$03 &  2.496$-$03 &  1.962$-$03 &  1.606$-$03 &  1.356$-$03 \\
  1 & 70 &  1.119$-$03 &  3.310$-$04 &  1.548$-$04 &  9.338$-$05 &  4.567$-$05 &  3.067$-$05 &  2.325$-$05 &  1.876$-$05 \\
  1 & 71 &  1.430$-$03 &  3.976$-$04 &  1.701$-$04 &  9.310$-$05 &  3.803$-$05 &  2.359$-$05 &  1.748$-$05 &  1.420$-$05 \\
  1 & 72 &  1.692$-$03 &  4.461$-$04 &  1.761$-$04 &  8.705$-$05 &  2.682$-$05 &  1.275$-$05 &  7.470$-$06 &  4.921$-$06 \\
  1 & 73 &  3.850$-$02 &  3.558$-$02 &  3.088$-$02 &  2.672$-$02 &  1.967$-$02 &  1.573$-$02 &  1.335$-$02 &  1.181$-$02 \\
  1 & 74 &  3.921$-$03 &  3.782$-$03 &  3.659$-$03 &  3.547$-$03 &  3.290$-$03 &  3.072$-$03 &  2.893$-$03 &  2.746$-$03 \\
  1 & 75 &  5.750$-$04 &  1.916$-$04 &  8.378$-$05 &  4.559$-$05 &  1.659$-$05 &  8.762$-$06 &  5.586$-$06 &  3.977$-$06 \\
  1 & 76 &  6.207$-$04 &  3.523$-$04 &  2.723$-$04 &  2.404$-$04 &  2.063$-$04 &  1.886$-$04 &  1.761$-$04 &  1.665$-$04 \\
  1 & 77 &  2.450$-$04 &  8.003$-$05 &  3.390$-$05 &  1.771$-$05 &  5.695$-$06 &  2.628$-$06 &  1.467$-$06 &  9.203$-$07 \\
  1 & 78 &  1.151$-$02 &  1.632$-$02 &  1.681$-$02 &  1.610$-$02 &  1.366$-$02 &  1.177$-$02 &  1.047$-$02 &  9.573$-$03 \\
  1 & 79 &  2.804$-$04 &  8.456$-$05 &  3.330$-$05 &  1.842$-$05 &  6.965$-$06 &  3.486$-$06 &  2.026$-$06 &  1.294$-$06 \\
  1 & 80 &  3.926$-$04 &  1.182$-$04 &  4.643$-$05 &  2.560$-$05 &  9.596$-$06 &  4.725$-$06 &  2.679$-$06 &  1.668$-$06 \\
  1 & 81 &  5.050$-$04 &  1.519$-$04 &  5.962$-$05 &  3.286$-$05 &  1.232$-$05 &  6.074$-$06 &  3.448$-$06 &  2.151$-$06 \\
  1 & 82 &  1.323$-$05 &  9.122$-$06 &  4.310$-$06 &  2.232$-$06 &  6.351$-$07 &  2.637$-$07 &  1.381$-$07 &  8.482$-$08 \\
  1 & 83 &  3.969$-$05 &  2.734$-$05 &  1.288$-$05 &  6.653$-$06 &  1.859$-$06 &  7.428$-$07 &  3.662$-$07 &  2.070$-$07 \\
  1 & 84 &  6.606$-$05 &  4.574$-$05 &  2.162$-$05 &  1.130$-$05 &  3.205$-$06 &  1.400$-$06 &  8.548$-$07 &  5.816$-$07 \\
  1 & 85 &  4.623$-$03 &  8.064$-$03 &  9.866$-$03 &  1.095$-$02 &  1.232$-$02 &  1.290$-$02 &  1.318$-$02 &  1.335$-$02 \\
  1 & 86 &  7.845$-$03 &  6.143$-$03 &  5.310$-$03 &  4.784$-$03 &  3.915$-$03 &  3.377$-$03 &  3.009$-$03 &  2.740$-$03 \\
  1 & 87 &  1.546$-$03 &  4.524$-$04 &  1.918$-$04 &  9.825$-$05 &  2.855$-$05 &  1.180$-$05 &  5.958$-$06 &  3.427$-$06 \\
  1 & 88 &  2.574$-$03 &  7.534$-$04 &  3.194$-$04 &  1.637$-$04 &  4.757$-$05 &  1.967$-$05 &  9.939$-$06 &  5.722$-$06 \\
  1 & 89 &  3.598$-$03 &  1.053$-$03 &  4.466$-$04 &  2.289$-$04 &  6.654$-$05 &  2.750$-$05 &  1.389$-$05 &  7.991$-$06 \\
  1 & 90 &  1.079$-$02 &  1.202$-$02 &  1.173$-$02 &  1.105$-$02 &  9.351$-$03 &  8.067$-$03 &  7.134$-$03 &  6.438$-$03 \\
  1 & 91 &  8.991$-$04 &  1.355$-$03 &  1.561$-$03 &  1.691$-$03 &  1.832$-$03 &  1.852$-$03 &  1.829$-$03 &  1.793$-$03 \\
  1 & 92 &  7.300$-$04 &  9.381$-$04 &  9.982$-$04 &  9.878$-$04 &  8.672$-$04 &  7.418$-$04 &  6.402$-$04 &  5.603$-$04 \\
  1 & 93 &  1.194$-$04 &  1.789$-$05 &  4.663$-$06 &  2.080$-$06 &  6.231$-$07 &  2.827$-$07 &  1.572$-$07 &  9.852$-$08 \\
  1 & 94 &  1.991$-$04 &  2.986$-$05 &  7.832$-$06 &  3.530$-$06 &  1.103$-$06 &  5.289$-$07 &  3.121$-$07 &  2.078$-$07 \\
  1 & 95 &  2.786$-$04 &  4.173$-$05 &  1.088$-$05 &  4.858$-$06 &  1.469$-$06 &  6.777$-$07 &  3.843$-$07 &  2.462$-$07 \\
  1 & 96 &  4.339$-$04 &  8.911$-$05 &  2.952$-$05 &  1.287$-$05 &  3.244$-$06 &  1.428$-$06 &  8.255$-$07 &  5.533$-$07 \\
  1 & 97 &  6.079$-$04 &  1.244$-$04 &  4.103$-$05 &  1.772$-$05 &  4.279$-$06 &  1.773$-$06 &  9.601$-$07 &  6.044$-$07 \\
  1 & 98 &  7.830$-$04 &  1.600$-$04 &  5.258$-$05 &  2.262$-$05 &  5.372$-$06 &  2.172$-$06 &  1.141$-$06 &  6.926$-$07 \\
  1 & 99 &  2.623$-$04 &  8.923$-$05 &  3.618$-$05 &  1.853$-$05 &  5.851$-$06 &  2.727$-$06 &  1.546$-$06 &  9.843$-$07 \\                                                                                                     
  1 &100 &  1.580$-$04 &  5.365$-$05 &  2.175$-$05 &  1.113$-$05 &  3.504$-$06 &  1.624$-$06 &  9.145$-$07 &  5.780$-$07 \\
  1 &101 &  5.276$-$05 &  1.789$-$05 &  7.248$-$06 &  3.708$-$06 &  1.166$-$06 &  5.398$-$07 &  3.035$-$07 &  1.914$-$07 \\
  1 &102 &  5.610$-$03 &  4.668$-$03 &  4.125$-$03 &  3.631$-$03 &  2.756$-$03 &  2.264$-$03 &  1.968$-$03 &  1.776$-$03 \\
  1 &103 &  1.817$-$03 &  2.277$-$03 &  2.214$-$03 &  2.058$-$03 &  1.733$-$03 &  1.538$-$03 &  1.420$-$03 &  1.345$-$03 \\
  1 &104 &  3.288$-$04 &  2.385$-$04 &  1.877$-$04 &  1.559$-$04 &  1.144$-$04 &  9.346$-$05 &  8.016$-$05 &  7.065$-$05 \\
  1 &105 &  2.992$-$05 &  8.670$-$06 &  3.722$-$06 &  1.991$-$06 &  6.639$-$07 &  3.091$-$07 &  1.714$-$07 &  1.060$-$07 \\
  1 &106 &  4.986$-$05 &  1.445$-$05 &  6.201$-$06 &  3.316$-$06 &  1.105$-$06 &  5.147$-$07 &  2.853$-$07 &  1.764$-$07 \\
  1 &107 &  6.979$-$05 &  2.022$-$05 &  8.678$-$06 &  4.639$-$06 &  1.546$-$06 &  7.198$-$07 &  3.989$-$07 &  2.465$-$07 \\
  1 &108 &  1.027$-$04 &  9.182$-$05 &  7.285$-$05 &  6.040$-$05 &  4.363$-$05 &  3.454$-$05 &  2.856$-$05 &  2.428$-$05 \\
  1 &109 &  1.255$-$04 &  1.856$-$05 &  6.572$-$06 &  3.231$-$06 &  9.764$-$07 &  4.336$-$07 &  2.329$-$07 &  1.409$-$07 \\
  1 &110 &  1.613$-$04 &  2.385$-$05 &  8.443$-$06 &  4.152$-$06 &  1.255$-$06 &  5.575$-$07 &  2.997$-$07 &  1.813$-$07 \\
  1 &111 &  1.970$-$04 &  2.911$-$05 &  1.030$-$05 &  5.068$-$06 &  1.532$-$06 &  6.808$-$07 &  3.658$-$07 &  2.213$-$07 \\
  1 &112 &  2.005$-$05 &  1.151$-$05 &  5.649$-$06 &  2.955$-$06 &  8.543$-$07 &  3.576$-$07 &  1.860$-$07 &  1.104$-$07 \\
  1 &113 &  3.799$-$04 &  9.867$-$05 &  2.974$-$05 &  1.388$-$05 &  4.427$-$06 &  2.039$-$06 &  1.110$-$06 &  6.712$-$07 \\
  1 &114 &  2.950$-$04 &  7.671$-$05 &  2.315$-$05 &  1.081$-$05 &  3.450$-$06 &  1.590$-$06 &  8.655$-$07 &  5.235$-$07 \\
  1 &115 &  2.105$-$04 &  5.479$-$05 &  1.655$-$05 &  7.740$-$06 &  2.476$-$06 &  1.143$-$06 &  6.243$-$07 &  3.789$-$07 \\
  1 &116 &  5.590$-$04 &  4.345$-$04 &  3.751$-$04 &  3.310$-$04 &  2.566$-$04 &  2.097$-$04 &  1.772$-$04 &  1.534$-$04 \\
  1 &117 &  1.981$-$04 &  3.630$-$05 &  1.171$-$05 &  5.234$-$06 &  1.395$-$06 &  5.892$-$07 &  3.104$-$07 &  1.865$-$07 \\
  1 &118 &  2.421$-$04 &  4.436$-$05 &  1.431$-$05 &  6.395$-$06 &  1.705$-$06 &  7.201$-$07 &  3.794$-$07 &  2.281$-$07 \\
  1 &119 &  2.861$-$04 &  5.242$-$05 &  1.691$-$05 &  7.555$-$06 &  2.014$-$06 &  8.507$-$07 &  4.483$-$07 &  2.695$-$07 \\
  1 &120 &  8.275$-$05 &  1.430$-$05 &  4.913$-$06 &  2.232$-$06 &  5.624$-$07 &  2.181$-$07 &  1.065$-$07 &  6.031$-$08 \\
  1 &121 &  1.158$-$04 &  2.001$-$05 &  6.874$-$06 &  3.123$-$06 &  7.860$-$07 &  3.042$-$07 &  1.482$-$07 &  8.357$-$08 \\
  1 &122 &  1.488$-$04 &  2.571$-$05 &  8.833$-$06 &  4.011$-$06 &  1.009$-$06 &  3.906$-$07 &  1.902$-$07 &  1.073$-$07 \\
  1 &123 &  6.740$-$05 &  1.048$-$04 &  1.101$-$04 &  1.051$-$04 &  8.714$-$05 &  7.172$-$05 &  6.006$-$05 &  5.134$-$05 \\
  1 &124 &  2.149$-$05 &  9.465$-$06 &  3.096$-$06 &  1.304$-$06 &  2.960$-$07 &  1.160$-$07 &  6.171$-$08 &  3.975$-$08 \\
  1 &125 &  6.445$-$05 &  2.837$-$05 &  9.265$-$06 &  3.886$-$06 &  8.574$-$07 &  3.160$-$07 &  1.524$-$07 &  8.587$-$08 \\
  1 &126 &  1.074$-$04 &  4.745$-$05 &  1.566$-$05 &  6.712$-$06 &  1.686$-$06 &  7.875$-$07 &  5.250$-$07 &  4.297$-$07 \\
  1 &127 &  9.344$-$04 &  1.290$-$03 &  1.482$-$03 &  1.599$-$03 &  1.752$-$03 &  1.818$-$03 &  1.849$-$03 &  1.866$-$03 \\
  1 &128 &  2.353$-$03 &  1.017$-$03 &  9.061$-$04 &  8.707$-$04 &  7.448$-$04 &  6.345$-$04 &  5.500$-$04 &  4.854$-$04 \\
  1 &129 &  7.508$-$05 &  9.825$-$06 &  3.443$-$06 &  1.824$-$06 &  5.949$-$07 &  2.625$-$07 &  1.376$-$07 &  8.073$-$08 \\
  1 &130 &  9.663$-$05 &  1.264$-$05 &  4.425$-$06 &  2.344$-$06 &  7.629$-$07 &  3.359$-$07 &  1.756$-$07 &  1.028$-$07 \\
  1 &131 &  1.183$-$04 &  1.546$-$05 &  5.408$-$06 &  2.863$-$06 &  9.319$-$07 &  4.102$-$07 &  2.145$-$07 &  1.255$-$07 \\
  1 &132 &  1.409$-$05 &  2.425$-$06 &  7.670$-$07 &  3.362$-$07 &  8.377$-$08 &  3.296$-$08 &  1.638$-$08 &  9.407$-$09 \\
  1 &133 &  2.347$-$05 &  4.042$-$06 &  1.277$-$06 &  5.592$-$07 &  1.387$-$07 &  5.413$-$08 &  2.658$-$08 &  1.500$-$08 \\
  1 &134 &  3.285$-$05 &  5.661$-$06 &  1.790$-$06 &  7.838$-$07 &  1.946$-$07 &  7.619$-$08 &  3.757$-$08 &  2.131$-$08 \\
  1 &135 &  4.677$-$05 &  1.140$-$05 &  4.392$-$06 &  2.259$-$06 &  8.514$-$07 &  4.619$-$07 &  2.889$-$07 &  1.960$-$07 \\
  1 &136 &  2.814$-$05 &  6.875$-$06 &  2.650$-$06 &  1.364$-$06 &  5.143$-$07 &  2.795$-$07 &  1.752$-$07 &  1.192$-$07 \\
  1 &137 &  9.392$-$06 &  2.296$-$06 &  8.850$-$07 &  4.552$-$07 &  1.713$-$07 &  9.287$-$08 &  5.804$-$08 &  3.935$-$08 \\
  1 &138 &  1.322$-$03 &  1.016$-$03 &  7.895$-$04 &  6.298$-$04 &  4.037$-$04 &  2.989$-$04 &  2.422$-$04 &  2.067$-$04 \\
  1 &139 &  9.768$-$04 &  4.668$-$04 &  3.533$-$04 &  2.909$-$04 &  2.009$-$04 &  1.525$-$04 &  1.228$-$04 &  1.030$-$04 \\
  1 &140 &  2.995$-$04 &  3.452$-$04 &  3.866$-$04 &  4.054$-$04 &  3.909$-$04 &  3.555$-$04 &  3.244$-$04 &  3.006$-$04 \\
  1 &141 &  5.470$-$04 &  1.071$-$03 &  1.346$-$03 &  1.555$-$03 &  1.954$-$03 &  2.242$-$03 &  2.454$-$03 &  2.618$-$03 \\
\hline
\end{longtable}

\clearpage 

\setlength{\LTleft}{0pt}
\setlength{\LTright}{0pt}

\renewcommand{\arraystretch}{1.1}

\footnotesize 

\begin{longtable}{@{\extracolsep\fill}rrrrrrrrrrrr@{}}
\caption{\label{tbl_tab4}
Effective collision strengths ($\Upsilon$) for  transitions of  Si III. $a{\pm}b \equiv$ $a\times$10$^{{\pm}b}$.
See page \pageref{tbl_tab1} for Explanation of Tables and Table~\ref{tbl_tab1} 
for definition of level indices.
}
Transition & & Temperature (log K) \\
\hline
$i$ &
$j$ &
4.10 & 4.30 & 4.50 & 4.70 & 4.90 & 5.10 & 5.30 & 5.50 & 5.70 & 5.90  \\
\hline
\endfirsthead\\
\caption[]{(continued)}
Transition & & Temperature (log K) \\
\hline
$i$ &
$j$ &
4.10 & 4.30 & 4.50 & 4.70 & 4.90 & 5.10 & 5.30 & 5.50 & 5.70 & 5.90  \\
\hline\\
\endhead
    1 &    2 &  5.720$-$01 &  4.905$-$01 &  4.174$-$01 &  3.474$-$01 &  2.798$-$01 &  2.176$-$01 &  1.643$-$01 &  1.212$-$01 &  8.748$-$02 &  6.194$-$02 \\
    1 &    3 &  1.765$+$00 &  1.501$+$00 &  1.270$+$00 &  1.053$+$00 &  8.461$-$01 &  6.569$-$01 &  4.952$-$01 &  3.645$-$01 &  2.629$-$01 &  1.862$-$01 \\
    1 &    4 &  2.899$+$00 &  2.465$+$00 &  2.089$+$00 &  1.736$+$00 &  1.397$+$00 &  1.086$+$00 &  8.195$-$01 &  6.034$-$01 &  4.353$-$01 &  3.080$-$01 \\
    1 &    5 &  5.925$+$00 &  6.415$+$00 &  6.924$+$00 &  7.528$+$00 &  8.472$+$00 &  9.919$+$00 &  1.172$+$01 &  1.372$+$01 &  1.598$+$01 &  1.862$+$01 \\
    1 &    6 &  7.875$-$01 &  8.377$-$01 &  8.773$-$01 &  9.052$-$01 &  9.363$-$01 &  9.879$-$01 &  1.067$+$00 &  1.173$+$00 &  1.298$+$00 &  1.431$+$00 \\
    1 &    7 &  2.222$-$02 &  1.975$-$02 &  1.658$-$02 &  1.321$-$02 &  1.006$-$02 &  7.372$-$03 &  5.245$-$03 &  3.650$-$03 &  2.497$-$03 &  1.683$-$03 \\
    1 &    8 &  6.780$-$02 &  5.993$-$02 &  5.012$-$02 &  3.985$-$02 &  3.029$-$02 &  2.221$-$02 &  1.584$-$02 &  1.104$-$02 &  7.555$-$03 &  5.085$-$03 \\
    1 &    9 &  1.159$-$01 &  1.018$-$01 &  8.490$-$02 &  6.738$-$02 &  5.117$-$02 &  3.748$-$02 &  2.671$-$02 &  1.867$-$02 &  1.289$-$02 &  8.838$-$03 \\
    1 &   10 &  4.833$-$01 &  4.649$-$01 &  4.363$-$01 &  3.984$-$01 &  3.541$-$01 &  3.064$-$01 &  2.576$-$01 &  2.094$-$01 &  1.642$-$01 &  1.241$-$01 \\
    1 &   11 &  3.444$-$01 &  3.313$-$01 &  3.109$-$01 &  2.838$-$01 &  2.522$-$01 &  2.183$-$01 &  1.835$-$01 &  1.492$-$01 &  1.170$-$01 &  8.847$-$02 \\
    1 &   12 &  2.061$-$01 &  1.981$-$01 &  1.858$-$01 &  1.697$-$01 &  1.509$-$01 &  1.309$-$01 &  1.103$-$01 &  8.979$-$02 &  7.048$-$02 &  5.330$-$02 \\
    1 &   13 &  3.283$-$01 &  2.970$-$01 &  2.551$-$01 &  2.103$-$01 &  1.676$-$01 &  1.300$-$01 &  9.830$-$02 &  7.254$-$02 &  5.226$-$02 &  3.679$-$02 \\
    1 &   14 &  2.307$-$01 &  2.282$-$01 &  2.247$-$01 &  2.238$-$01 &  2.285$-$01 &  2.393$-$01 &  2.546$-$01 &  2.726$-$01 &  2.908$-$01 &  3.062$-$01 \\
    1 &   15 &  4.601$-$01 &  4.514$-$01 &  4.465$-$01 &  4.529$-$01 &  4.787$-$01 &  5.162$-$01 &  5.503$-$01 &  5.768$-$01 &  6.003$-$01 &  6.250$-$01 \\
    1 &   16 &  7.080$-$01 &  7.267$-$01 &  7.455$-$01 &  7.731$-$01 &  8.182$-$01 &  8.812$-$01 &  9.567$-$01 &  1.039$+$00 &  1.123$+$00 &  1.202$+$00 \\
    1 &   17 &  2.898$-$02 &  2.824$-$02 &  2.627$-$02 &  2.345$-$02 &  2.025$-$02 &  1.694$-$02 &  1.371$-$02 &  1.072$-$02 &  8.107$-$03 &  5.940$-$03 \\
    1 &   18 &  8.786$-$02 &  8.557$-$02 &  7.953$-$02 &  7.079$-$02 &  6.076$-$02 &  5.051$-$02 &  4.070$-$02 &  3.179$-$02 &  2.406$-$02 &  1.769$-$02 \\
    1 &   19 &  1.494$-$01 &  1.453$-$01 &  1.348$-$01 &  1.197$-$01 &  1.025$-$01 &  8.494$-$02 &  6.826$-$02 &  5.315$-$02 &  4.011$-$02 &  2.936$-$02 \\
    1 &   20 &  2.091$-$01 &  2.170$-$01 &  2.220$-$01 &  2.272$-$01 &  2.355$-$01 &  2.482$-$01 &  2.649$-$01 &  2.848$-$01 &  3.069$-$01 &  3.306$-$01 \\
    1 &   21 &  5.064$-$02 &  4.433$-$02 &  3.876$-$02 &  3.382$-$02 &  2.918$-$02 &  2.463$-$02 &  2.014$-$02 &  1.587$-$02 &  1.203$-$02 &  8.790$-$03 \\
    1 &   22 &  7.475$-$02 &  6.484$-$02 &  5.630$-$02 &  4.892$-$02 &  4.211$-$02 &  3.546$-$02 &  2.893$-$02 &  2.274$-$02 &  1.720$-$02 &  1.254$-$02 \\
    1 &   23 &  9.696$-$02 &  8.375$-$02 &  7.273$-$02 &  6.344$-$02 &  5.487$-$02 &  4.636$-$02 &  3.787$-$02 &  2.976$-$02 &  2.249$-$02 &  1.638$-$02 \\
    1 &   24 &  5.804$-$02 &  5.532$-$02 &  5.238$-$02 &  4.895$-$02 &  4.472$-$02 &  3.961$-$02 &  3.377$-$02 &  2.764$-$02 &  2.169$-$02 &  1.636$-$02 \\
    1 &   25 &  9.736$-$02 &  9.262$-$02 &  8.759$-$02 &  8.176$-$02 &  7.464$-$02 &  6.605$-$02 &  5.630$-$02 &  4.606$-$02 &  3.615$-$02 &  2.727$-$02 \\
\hline
\end{longtable}


\section*{References}
\begin{thebibliography}{999}
\bibitem[\protect\citeauthoryear{Dufton \& Kingston}{1989}]{dk1} 
P.L. Dufton, A.E. Kingston. Month. Not. R. Astron. Soc. 241 (1989) 209.
\bibitem[\protect\citeauthoryear{Fern{\'a}ndez-Menchero, Del Zanna \& Badnell}{Fern{\'a}ndez-Menchero et al.}{2014}]{icft1}
L.  Fern{\'a}ndez-Menchero, G. Del Zanna,  N.R. Badnell.  Astron. Astrophys.  572 (2014) A115. 
\bibitem[\protect\citeauthoryear{Badnell}{2011}]{as}
N.R. Badnell. Comput. Phys. Commun.  182 (2011)  1528.
\bibitem[\protect\citeauthoryear{Berrington, Eissner \& Norrington}{Berrington et al.}{1995}]{rm2}
K.A. Berrington, W.B. Eissner, P.H. Norrington P. H.  Comput. Phys. Commun.  92 (1995)  290.
\bibitem[\protect\citeauthoryear{Griffin, Badnell \& Pindzola}{Griffin  et al.}{1998}]{icft}
D.C. Griffin, N.R. Badnell,  M.S. Pindzola.  J. Phys. B 31 (1998) 3713.
\bibitem[\protect\citeauthoryear{Aggarwal \& Keenan}{2015}]{alx1}
K.M. Aggarwal,   F.P. Keenan.  Month. Not. R. Astron. Soc. 447 (2015) 3849. 
\bibitem[\protect\citeauthoryear{Aggarwal \& Keenan}{2015}]{ciii}
K.M. Aggarwal,   F.P. Keenan. Month. Not. R. Astron. Soc. 450 (2015) 1151.
\bibitem[\protect\citeauthoryear{Aggarwal \& Keenan}{2014b}]{fe14}
K.M. Aggarwal,   F.P. Keenan. Month. Not. R. Astron. Soc. 445 (2014) 2015.
\bibitem[\protect\citeauthoryear{Tayal \& Zatsarriny}{2015}]{fe9}
S.S. Tayal, O. Zatsarriny. Astrophys. J.  812 (2015) 174.
\bibitem[\protect\citeauthoryear{Fern{\'a}ndez-Menchero, Del Zanna \& Badnell}{Fern{\'a}ndez-Menchero et al.}{2015}]{icft2}
 L. Fern{\'a}ndez-Menchero, G.  Del Zanna,  N.R. Badnell. Month. Not. R. Astron. Soc. 450 (2015) 4174.
\bibitem[\protect\citeauthoryear{Del Zanna  et al.}{2015}]{icft3}
G.  Del Zanna,  N.R. Badnell, L. Fern{\'a}ndez-Menchero, G.Y. Liang, H.E. Mason, P.J. Storey. Month. Not. R. Astron. Soc. 454 (2015) 2909.
\bibitem[\protect\citeauthoryear{Badnell  et al.}{2016}]{icft4}
N.R. Badnell, G. Del Zanna, L. Fern{\'a}ndez-Menchero, A.S. Giunta, G.Y. Liang, H.E.  Mason, P.J. Stroey.  J. Phys. B 49 (2016) 094001.
\bibitem[\protect\citeauthoryear{Aggarwal \& Keenan}{2015}]{ti19}
K.M. Aggarwal,   F.P. Keenan.  Phys. Scr. 86  (2012) 055301. 
\bibitem[\protect\citeauthoryear{Aggarwal \& Keenan}{2015}]{cl14}
K.M. Aggarwal,   F.P. Keenan.  Phys. Scr. 89  (2014) 125401. 
\bibitem[\protect\citeauthoryear{Aggarwal \& Keenan}{2015}]{alx}
K.M. Aggarwal,   F.P. Keenan.  Month. Not. R. Astron. Soc. 438 (2014) 1223. 
 \bibitem[\protect\citeauthoryear{Fern{\'a}ndez-Menchero, Del Zanna \& Badnell}{Fern{\'a}ndez-Menchero et al.}{2014}]{belike}
 L. Fern{\'a}ndez-Menchero, G. Del Zanna,  N.R. Badnell N.R.  Astron. Astrophys. 566, (2014) A104. 
\bibitem[\protect\citeauthoryear{Aggarwal, Keenan \& Lawson}{Aggarwal et al.}{2016}]{niv}
K.M. Aggarwal,   F.P. Keenan, K.D. Lawson K. D.  Month. Not. R. Astron. Soc. 461 (2016) 3997.
\bibitem[\protect\citeauthoryear{Griffin  et al.}{1999}]{icft5}
D.C. Griffin, N.R. Badnell, M.S. Pindzola, J.A.  Shaw. J. Phys.  B 32 (1999) 2139.
\bibitem[\protect\citeauthoryear{Aggarwal et al.}{2000}]{fe15}
K.M. Aggarwal,  N.C.  Deb, F.P. Keenan, A.Z. Msezane. J. Phys.  B 33 (2000)  L391.
\bibitem[\protect\citeauthoryear{Berrington et al.}{2005}]{icft6}
K.A. Berrington, C.P. Ballance, D.C. Griffin, N.R. Badnell.  J. Phys. B 38 (2005) 1667.
\bibitem[\protect\citeauthoryear{Storey, Sochi \& Badnell}{Storey et al.}{2014}]{oiii}
P.J. Storey, T. Sochi, N.R.  Badnell. Month. Not. R. Astron. Soc. 441 (2014) 3028.
\bibitem[\protect\citeauthoryear{Del Zanna, Fern{\'a}ndez-Menchero \& Badnell}{Del Zanna  et al.}{2015}]{bench}
 G. Del Zanna, L.  Fern{\'a}ndez-Menchero, N.R. Badnell.  Astron. Astrophys.  574 (2015) A99.
\bibitem[\protect\citeauthoryear{Linsky et al.}{1995}]{jll}
J.L. Linsky, B.E. Wood, P. Judge,A.  Brown, C. Andrulis, T.R. Ayers.  Astrophys.  J. 442 (2009) 381. 
\bibitem[\protect\citeauthoryear{Baluja, Burke \& Kingston}{Baluja et al.}{1981}]{bbk}
K.L. Baluja, P.G. Burke, A.E. Kingston.  J. Phys.  B 14 (1981)  1333. 
\bibitem[\protect\citeauthoryear{Bryans, Landi \& Savin}{Bryans et al.}{2009}]{pb}
P. Bryans, E. Landi,   D.W.  Savin. Astrophys. J.  691 (2009) 1540.  
\bibitem[\protect\citeauthoryear{Baldwin et al.}{1996}]{bald}
J.A. Baldwin, G.J. Ferland, K.T. Korista, R.F. Carswell, F. Hamann, M.M. Phillips, D. Verner, B.J. Wilkes, R.E. Williams.  Astrophys. J.  461 (1996) 664. 
\bibitem[\protect\citeauthoryear{Grant et al.}{1980}]{grasp0}
I.P. Grant, B.J. McKenzie,  P.H. Norrington,  D.F. Mayers,   N.C. Pyper N. C.    Comput. Phys. Commun.  21 (1980)  207.  
\bibitem[\protect\citeauthoryear{Aggarwal \& Keenan}{2014}]{si2}
K.M. Aggarwal,   F.P. Keenan F. P. Month. Not. R. Astron. Soc. 442 (2014) 388.   
\bibitem{fac}
M.F. Gu,  Can. J.  Phys, 86 (2008) 675.
\bibitem[\protect\citeauthoryear{Safronova, Johnson \& Berry}{Safronova et al.}{2000}]{saf}
U.I. Safronova, W.R. Johnson, H.G. Berry. Phys. Rev. A 61 (2000) 052503.
\bibitem[\protect\citeauthoryear{Kelleher \& Podobedova}{2008}]{kp} 
D.E. Kelleher, L.I. Podobedova. J. Phys. Chem. Ref. Data 37 (2008) 1285.
 \bibitem[\protect\citeauthoryear{Froese Fischer, Tachiev \& Irimia}{Froese Fischer et al.}{2006}]{cff} 
C.  Froese Fischer, TG. achiev, A. Irimia A. At. Data Nucl. Data Tables 92 (2006) 607.
 \bibitem{co16}   
K.M. Aggarwal, V. Tayal, G.P. Gupta, F.P. Keenan,  At. Data Nucl. Data Tables  93 (2007) 615.
\bibitem[\protect\citeauthoryear{Berry et al.}{1971}]{berry}
H.G. Berry, J. Bromander, L.J. Curtis, R. Buchta R. Phys. Scr. 3 (1971) 125.
\bibitem[\protect\citeauthoryear{Bashkin et al.}{1980}]{bash}
S. Bashkin,  G. Astner, S. Mannervik, P.S. Ramanujam, M. Scofield, S. Huldt, I. Martinson. Phys. Scr. 21 (1980) 820.
\bibitem[\protect\citeauthoryear{Kwong et al.}{1983}]{hsk}
H.S. Kwong, B.C. Johnson, P.L. Smith, W.H. Parkinson. Phys. Rev.,A 27 (1983) 3040.
\bibitem[\protect\citeauthoryear{Livingston et al.}{1976}]{liva}
A.E. Livingston, Y. Baudinet-Robinet, H.P. Garnir, P.D. Dumont. J. Opt. Soc. Am. 66 (1976) 1393.
\bibitem[\protect\citeauthoryear{Livingston et al.}{1976}]{livb}
A.E. Livingston, J.A. Kernahan, D.J.G. Irwin, E.H. Pinnington.  J. Phys. B 9 (1976) 389.
\bibitem[\protect\citeauthoryear{Burgess \& Sheorey}{1974}]{ab}
A. Burgess,   V.B. Sheorey.  J. Phys.  B7 (1974)  2403.
\bibitem[\protect\citeauthoryear{Wallbank et al.}{1997}]{wall}
B. Wallbank, N. Djuri{\'c}, O. Woitk., S. Zhou, G.H. Dunn, A.C.H. Smith, M.E. Bannister. Phys. Rev. A 56 (1997) 3714.
\bibitem[\protect\citeauthoryear{Risenfeld et al.}{1999}]{rise}
D.B. Reisenfeld, L.D. Gardner, P.H. Janzen, D.W. Savin, J.L. Kohl.  Phys. Rev. A 60 (1999) 1153.
\bibitem[\protect\citeauthoryear{Kai, Srivastava \& Nakazaki}{Kai  et al.}{2004}]{kai}
T.  Kai, R. Srivastava, S. Nakazaki  J. Phys. B 37 (2004) 2045.
\bibitem[\protect\citeauthoryear{Griffin, Pindzola \& Badnell}{Griffin  et al.}{1993}]{dcg}
D.C. Griffin, M.S. Pindzola, N.R. Badnell, Phys. Rev.  A 47 (1993) 2871.
 

\end{thebibliography}
\end{document}